\pdfoutput=1

\documentclass[11pt,twoside,a4paper,cmspaper,final,collab]{cms-tdr}
\def\svnVersion{464987P}\def\cmsCernNoTag{CERN-EP-2017-244}\def\cmsCernDate{\today}\def\cmsMessage{Published in the European Physical Journal C as \href{http://dx.doi.org/10.1140/epjc/s10052-018-5929-3}{\doi{10.1140/epjc/s10052-018-5929-3.}}}

\begin{document}\cmsNoteHeader{BPH-13-008}

\hyphenation{had-ron-i-za-tion}
\hyphenation{cal-or-i-me-ter}
\hyphenation{de-vices}

\RCS$Revision: 464497 $
\RCS$HeadURL: svn+ssh://svn.cern.ch/reps/tdr2/papers/BPH-13-008/trunk/BPH-13-008.tex $
\RCS$Id: BPH-13-008.tex 464497 2018-06-14 07:43:40Z jmejiagu $

\cmsNoteHeader{BPH-13-008}

\newcommand{\Jpsi}{\ensuremath{\cPJgy}}
\newcommand{\Jpsinarrow}{\ensuremath{\cmsSymbolFace{J}\hspace{-.14em}/\hspace{-.20em}\psi}}
\newcommand{\PKstnarrow}{\ensuremath{\mathrm{K^{\hspace{-0.02em}\ast}\hspace{-0.10em}(892)}}}
\newcommand{\PbgLp}{\ensuremath{\Lambda_\mathrm{b}^0}}
\newcommand{\Bs}{\ensuremath{\PBzs}}
\newcommand{\Bd}{\ensuremath{\PBz}}
\newcommand{\Bu}{\ensuremath{\PBp}}
\newcommand{\Bq}{\ensuremath{\mathrm{B}_{\mathrm{q}}}}
\newcommand{\Kstar}{\ensuremath{\PKst{}^{0}}}
\newcommand{\Lamb}{\ensuremath{\PgL^{0}}}
\newcommand{\BsJpsiKK}{\ensuremath{\PBzs \!\to\! \cPJgy \PKp \PKm}}
\newcommand{\BpJpsiKp}{\ensuremath{\PBp \!\to\! \cPJgy \PKp}}
\newcommand{\BsJpsiPhi}{\ensuremath{\PBzs \!\to\! \cPJgy \Pgf}}
\newcommand{\BdJpsiKstar}{\ensuremath{\PBz \!\to\! \cPJgy \PKst{}^{0}}}
\newcommand{\BdJpsiKshort}{\ensuremath{\PBz \!\to\! \cPJgy \PKzS}}
\newcommand{\BsJpsiPiPi}{\ensuremath{\PBzs \!\to\! \cPJgy \Pgpp \Pgpm}}
\newcommand{\BdJpsiPiPi}{\ensuremath{\PBz \!\to\! \cPJgy \Pgpp \Pgpm}}
\newcommand{\LambJpsiLam}{\ensuremath{\PbgLp \!\to\!\cPJgy \PgL^{0}}}
\newcommand{\JpsiPhi}{\ensuremath{\cPJgy \Pgf}}
\newcommand{\JpsiPiPi}{\ensuremath{\cPJgy \Pgpp \Pgpm}}
\newcommand{\JpsiKshort}{\ensuremath{ \cPJgy \PKzS}}
\newcommand{\JpsiKstar}{\ensuremath{\cPJgy \PKst{}^{0}}}
\newcommand{\JpsiLam}{\ensuremath{\cPJgy \PgL^{0}}}
\newcommand{\ct}          {\ensuremath{ct}\xspace}
\newcommand{\PBc}         {\ensuremath{\mathrm{B}_\mathrm{c}^+}\xspace}
\newcommand{\BcJpsiPip}   {\ensuremath{\PBc\!\to\!\cPJgy \Pgpp}\xspace}
\newcommand{\BcJpsiKp}    {\ensuremath{\PBc\!\to\!\cPJgy \PKp}\xspace}
\newcommand{\JpsiPi}     {\ensuremath{\cPJgy \Pgpp}\xspace}
\newcommand{\JpsiK}      {\ensuremath{\cPJgy \PKp}\xspace}
\newcommand{\BJpsiKp}     {\ensuremath{\PBp\!\to\!\cPJgy \PKp}\xspace}
\newcommand{\PaBzs}{\ensuremath{\overline{\cmsSymbolFace{B}}{}^0_\cmsSymbolFace{s}}\xspace}

\title{Measurement of b hadron lifetimes in pp collisions at $\sqrt{s} = 8$\TeV}

\date{\today}

\abstract{
Measurements are presented of the lifetimes of the \PBz, \PBzs, $\PgL_\PQb^0$, and \PBc hadrons using the decay channels
\BdJpsiKstar, \BdJpsiKshort, \BsJpsiPiPi, \BsJpsiPhi, \LambJpsiLam,  and \BcJpsiPip.
The data sample, corresponding to an integrated luminosity of 19.7\fbinv, was collected by the CMS detector at the LHC in proton-proton collisions at $\sqrt{s}=8\TeV$.
The \PBz\ lifetime
is measured to be $453.0 \pm 1.6\stat \pm 1.8\syst\micron$ in \JpsiKstar and
$457.8 \pm 2.7\stat \pm 2.8\syst\micron$ in \JpsiKshort, which results in a combined measurement of
$c\tau_{\PBz} = 454.1 \pm 1.4\stat \pm 1.7\syst\micron$.
The effective lifetime of the \PBzs meson is measured in two decay modes, with contributions from different amounts of the heavy and light
eigenstates.  This results in two different measured lifetimes: $c\tau_{\PBzs \to \cPJgy \Pgpp \Pgpm} = 502.7 \pm 10.2\stat\pm 3.4\syst\micron$ and $c\tau_{\PBzs \to \cPJgy \Pgf} = 443.9 \pm 2.0\stat \pm1.5\syst\micron$. The $\PgL_\PQb^0$ lifetime
is found to be $442.9 \pm 8.2\stat \pm 2.8\syst\micron$.
The precision from each of these channels is as good as or
better than previous measurements.
The \PBc lifetime, measured with respect to the \PBp\ to reduce the systematic uncertainty, is
$162.3 \pm 7.8\stat \pm 4.2\syst \pm 0.1\,(\tau_{\PBp})\micron$.
All results are in agreement with current world-average values.
}

\hypersetup{%
pdfauthor={CMS Collaboration},%
pdftitle={Measurement of b hadron lifetimes in pp collisions at sqrt(s) = 8 TeV},%
pdfsubject={CMS},%
pdfkeywords={b hadrons, lifetimes}}

\maketitle

\section{Introduction}

Precise lifetime measurements involving the weak interaction play an important role in the study of nonperturbative aspects of quantum chromodynamics (QCD). The phenomenology is commonly described by the QCD-inspired heavy-quark expansion model, which provides estimates of the ratio of lifetimes for hadrons containing a common heavy quark~\cite{Lenz:2015dra}. In this paper, we report measurements of the lifetimes of the \Bd{}, \Bs{}, \PbgLp{}, and \PBc hadrons.

The measurements are based on the reconstruction of the transverse decay length $L_{xy}$, where $\vec{L}_{xy}$ is defined as the flight distance
vector from the primary vertex to the decay vertex of the \cPqb{} hadron, projected onto the transverse component $\vec{p}_\text{T}$ (perpendicular to the beam axis) of the \cPqb{} hadron momentum. The proper decay time of the \cPqb{} hadron times the speed of light is measured using
\begin{equation}
ct =c L_{xy}\frac{M}{\pt},
\label{eq:ct_formula}
\end{equation}
where $M$ is the world-average value of the mass of the \cPqb{} hadron~\cite{PDG_lifetimes16}.

{\tolerance=900
In this analysis, the \cPqb{} hadrons are reconstructed from decays containing a \Jpsi{} meson. The data were recorded by the CMS detector~\cite{Chatrchyan:2008zzk} at the CERN LHC using dedicated triggers that require two oppositely charged muons consistent with originating from a common vertex and with an invariant mass compatible with that of the \Jpsi{} meson. Specifically, we reconstruct the decay modes \BdJpsiKstar, \BdJpsiKshort, \BsJpsiPiPi, \BsJpsiPhi, \LambJpsiLam, and \BcJpsiPip, where $\Jpsi \!\to\! \mu^+ \mu^-$, $\PKst{}^{0} \!\to\! \PKp \Pgpm$, $\PKzS \!\to\! \Pgpp \Pgpm$, $\Pgf \!\to\! \PKp \PKm$, and $\Lamb \!\to\! \Pp \Pgpm$. The \BJpsiKp decay is used as a reference mode and in evaluating some of the systematic uncertainties. Charge conjugation is implied throughout, unless otherwise indicated.
\par}

The decay rate of neutral \Bq{} (q = s or d) mesons is characterized by two parameters: the average decay width $\Gamma_{\mathrm{q}} = (\Gamma_{\mathrm{L}}^{\mathrm{q}} + \Gamma_{\mathrm{H}}^{\mathrm{q}})/2$ and the decay width difference $\Delta\Gamma_{\mathrm{q}} = \Gamma_{\mathrm{L}}^{\mathrm{q}} - \Gamma_{\mathrm{H}}^{\mathrm{q}}$, where $\Gamma_{\mathrm{L,H}}^{\mathrm{q}}$ are the decay widths of the light (L) and heavy (H) mass eigenstates. Assuming equal amounts of \Bq{} and its antiparticle are produced in the proton-proton collisions, the time-dependent decay rate into a final state $f$ that is accessible by both particle and antiparticle can be written as~\cite{Fleischer:2011cw}:
\begin{equation}
 R_{\mathrm{L}}^{f} \re^{- \Gamma_{\mathrm{L}}^{\mathrm{q}} t } +   R_{\mathrm{H}}^{f} \re^{- \Gamma_{\mathrm{H}}^{\mathrm{q}} t},
\end{equation}
where $R_{\mathrm{L}}^{f}$ and $R_{\mathrm{H}}^{f}$ are the amplitudes of the light and heavy mass states, respectively. Since the neutral B mesons have two eigenstates with different lifetimes, the \ct distribution consists of the sum of two exponential contributions. The effective lifetime of the neutral \Bq{} meson, produced as an equal admixture of particle and antiparticle flavour eigenstates and decaying into a final state $f$, can be written as~\cite{Fleischer:2011cw}:
\begin{equation}
\tau_\text{eff} = \frac{ \frac{ R_\mathrm{L}^f}{ \left(\Gamma_{\mathrm{L}}^{\mathrm{q}} \right)^2 } + \frac{R_\mathrm{H}^f}{ \left(\Gamma_\mathrm{H}^{\mathrm{q}} \right)^2 }}{  \frac{R_\mathrm{L}^f}{\Gamma_{\mathrm{L}}^{\mathrm{q}} } + \frac{R_\mathrm{H}^f}{\Gamma_{\mathrm{H}}^{\mathrm{q}} } }.
 \label{eq:effective_tau}	
\end{equation}
Since the amplitudes $R_{\mathrm{H}}^f$ and $R_{\mathrm{L}}^f$  are specific to the decay channel, the effective lifetime depends on the final state $f$ and is measured by fitting an exponential function to a distribution consisting of the sum of two exponential contributions. Because the \Bd{} system has a small lifetime difference with respect to the average lifetime, $\Delta \Gamma_\mathrm{d}/\Gamma_\mathrm{d} = (-0.2 \pm 1.0)\%$~\cite{Amhis:2016xyh}, the \ct distribution is close to an exponential, and it is treated as such for the lifetime measurement.  Following Ref.~\cite{DGtheory}, the \Bd{} lifetimes measured in the flavour-specific channel \BdJpsiKstar\ and the $CP$ eigenstate channel \BdJpsiKshort\ are used to determine values for $\Delta \Gamma_\mathrm{d}$, $\Gamma_\mathrm{d}$, and $\Delta \Gamma_\mathrm{d}/\Gamma_\mathrm{d}$.

{\tolerance=900
In the \Bs{} system, $\Delta \Gamma_\mathrm{s}/\Gamma_{\mathrm{s}} = (13.0 \pm 0.9)\%$~\cite{Amhis:2016xyh} and the deviation from an exponential \ct distribution is sizeable.  In this analysis, the two lifetimes associated with the \Bs{} meson are measured in the \JpsiPiPi{} and \JpsiPhi{} decay channels. The \BsJpsiPiPi{} decays are reconstructed in the invariant mass range $0.9240 < M( \pi^{+}\pi^{-} ) < 1.0204$\GeV, which is dominated by the $f_{0}(980)$ resonance~\cite{lhcbf01,lhcbf02}, making it a CP-odd final state.  Therefore, the lifetime measured in this channel is related to the inverse of the decay width of the heavy \Bs{} mass eigenstate, $\tau_{\Bs}^{\text{CP-odd}} \approx 1/\Gamma_\mathrm{H}$, as CP violation in mixing is measured to be negligible~\cite{PDG_lifetimes16}. The \JpsiPhi{} decay channel is an admixture of CP-even and CP-odd states, corresponding to the light and heavy mass eigenstates, respectively, neglecting CP violation in mixing. Rewriting Eq.~(\ref{eq:effective_tau}), the effective lifetime of the \Bs{} meson decaying to \JpsiPhi{} can be expressed as
\begin{equation}
 	\tau_{\text{eff}} = f_\mathrm{H} \tau_\mathrm{H} + (1-f_\mathrm{H}) \tau_\mathrm{L},
 \label{eq:Bs_tau_eff}	
\end{equation}
where $\tau_\mathrm{L}$ and $\tau_\mathrm{H}$ are the lifetimes of the light and heavy mass states, respectively, and $f_\mathrm{H}$ is the
heavy-component fraction, defined as:
\begin{equation}
f_\mathrm{H} = \frac{| A_{\perp} |^2 \tau_\mathrm{H}}{ |A|^2 \tau_\mathrm{L} +  |A_\perp |^2 \tau_\mathrm{H}}.
\end{equation}
Here, $|A|^2 = | A_0(0) |^2 + | A_{\parallel}(0) |^2$ is the sum of the squares of the amplitudes of the two CP-even states, and $| A_{\perp} |^2 = | A_{\perp}(0) |^2$ is the square of the amplitude of the CP-odd state. The amplitudes are determined at the production time $t=0$. Normalization constraints require $|A|^2 = 1 - |A_\perp|^2$ and therefore
\begin{equation}
f_\mathrm{H} = \frac{| A_{\perp} |^2 \tau_\mathrm{H}}{ (1 - |A_\perp|^2) \tau_\mathrm{L} +  |A_\perp |^2 \tau_\mathrm{H}}.
\end{equation}
By combining the \Bs{} lifetimes obtained from the final states \JpsiPhi{} and \JpsiPiPi{}, it is possible to determine the lifetime of the light \Bs{} mass eigenstate. The results in this paper are complementary to the CMS weak mixing phase analysis in the \BsJpsiPhi{} channel~\cite{phisPaper}, which provided measurements of the average decay width $\Gamma_\mathrm{s}$ and the decay width difference $\Delta \Gamma_\mathrm{s}$.
\par}

The weak decay of the \PBc meson can occur through either the b or c quark decaying, with the other quark as a spectator, or through an annihilation process.
The latter is predicted to contribute 10\% of the decay width~\cite{Kiselev:2002vz}, and lifetime measurements can be used to test the \PBc decay model.
As fewer and less precise measurements of the \PBc lifetime exist~\cite{BcLifetimeLHCB,BcLifetimeLHCBhad,CDF_lf_1998,CDF_lf,D0_lf,CDF_lf_hadr} compared to other b hadrons, the \PBc lifetime measurement presented in this paper is particularly valuable.

\section{The CMS detector}

The central feature of the CMS apparatus is a superconducting solenoid of 6\unit{m} internal diameter, providing a
magnetic field of 3.8\unit{T}. Within the solenoid volume are a silicon pixel and strip tracker, a lead tungstate crystal
electromagnetic calorimeter, and a brass and scintillator hadron calorimeter, each composed of a barrel and two endcap sections.
Forward calorimeters extend the pseudorapidity coverage provided by the barrel and endcap detectors. Muons are detected in
gas-ionization chambers embedded in the steel flux-return yoke outside the solenoid.

The main subdetectors used for this analysis are the silicon tracker and the muon detection system.
The silicon tracker measures charged particles in the pseudorapidity range $\abs{\eta}<2.5$. It consists
of 1440 silicon pixel and 15\,148 silicon strip detector modules. For charged particles of $1 < \pt < 10\GeV$ and $\abs{\eta} < 1.4$,
the track resolutions are typically 1.5\% in \pt and 25--90 (45--150)\mum in the transverse (longitudinal) impact parameter~\cite{TRK-11-001}. Muons are measured in the pseudorapidity range $\abs{\eta} < 2.4$, with detection planes made using three technologies: drift tubes,
cathode strip chambers, and resistive-plate chambers.

Events of interest are selected using a two-tiered trigger system~\cite{Khachatryan:2016bia}. The first level,
composed of custom hardware processors, uses information from the calorimeters and muon detectors to select events at a
rate of around 100\unit{kHz} within a time interval of less than 4\mus. The second level, known as the high-level
trigger (HLT), consists of a farm of processors running a version of the full event reconstruction software optimized
for fast processing, and reduces the event rate to around 1\unit{kHz} before data storage.  At the HLT stage, there is
full access to the event information, and therefore selection criteria similar to those applied
offline can be used.

A more detailed description of the CMS detector, together with a definition of the coordinate system used and the relevant kinematic variables,
can be found in Ref.~\cite{Chatrchyan:2008zzk}.

\section{Data and Monte Carlo simulated samples}

The data used in this analysis were collected in 2012 from proton-proton collisions at a centre-of-mass energy of 8\TeV, and correspond to an integrated luminosity of 19.7\fbinv.

{\tolerance=900
Fully simulated Monte Carlo (MC) samples of \BpJpsiKp, \BdJpsiKstar, \BdJpsiKshort, \BsJpsiPiPi, \BsJpsiPhi, and \LambJpsiLam{} were produced with
\PYTHIA 6.424~\cite{pythia} to simulate the proton-proton collisions, and subsequent parton shower and hadronization processes. The \PBc MC sample was produced with the dedicated generator \textsc{bcvegpy} 2.0~\cite{Bcveg1,Bcveg2} interfaced to \PYTHIA{}.
Decays of particles containing b or c quarks are simulated with the \EVTGEN{} package~\cite{evt}, and final-state radiation is included via \PHOTOS{}~\cite{Photos}.
Events are passed through the CMS detector simulation based on \GEANTfour{}\cite{geant4}, including additional proton-proton collisions in the same or
nearby beam crossings (pileup) to match the number of multiple vertices per event in the data. Simulated events are processed with the same reconstruction and trigger algorithms as the data.
\par}

\section{Reconstruction of \cPqb{} hadrons}\label{sec:recob}

The data are collected with a trigger that is designed to identify events in which a \Jpsi{} meson decays to two oppositely charged muons. The transverse momentum of the \Jpsi{} candidate is required to be greater than 7.9\GeV and both muons must be in the pseudorapidity region $|\eta| < 2.2$.  The distance of closest approach of each muon to the event vertex in the transverse plane must be less than 0.5\cm and a fit of the two muons to a common vertex must have a $\chi^2$ probability greater than 0.5\%. The invariant mass of the dimuon system must lie within $\pm$5 times the experimental  mass resolution (typically about 35\MeV) of the world-average \Jpsi{} mass~\cite{PDG_lifetimes16}.

The offline selection starts from \Jpsi{} candidates that are reconstructed from pairs of oppositely charged muons.
The standard CMS muon reconstruction procedure~\cite{Chatrchyan:2012xi} is used to identify the muons, which requires multiple hits in the pixel, strip, and muon detectors with a consistent trajectory throughout.
The offline selection requirements on the dimuon system replicate the trigger selection. From the sample of collected \Jpsi{} events, candidate b hadrons are reconstructed by combining a \Jpsi{} candidate with track(s) or reconstructed neutral particles, depending on the decay mode. Only tracks that pass the standard CMS high-purity requirements~\cite{TRK-11-001} are used. The b hadron candidate is fitted to a common vertex with the appropriate masses assigned to the charged tracks and the dimuon invariant mass constrained to the world-average \Jpsi{} mass~\cite{PDG_lifetimes16}. In fits that include a \PKzS{} or $\PgL^0$ hadron, the world-average mass is used for those particles. Primary vertices (PV) are fitted from the reconstructed tracks using an estimate of the proton-proton interaction region (beamspot) as a constraint. The PV having the smallest pointing angle, defined as the angle between the reconstructed \cPqb{} hadron momentum and the vector joining the PV with the decay vertex, is used.  As the proper decay times are measured in the transverse plane, where the PV position is dominated by the beamspot, the choice of PV has little effect on the analysis and is accounted for as a systematic uncertainty.

\subsection{Reconstruction of \texorpdfstring{\Bu{}}{B+}, \texorpdfstring{\Bd{}}{B0}, \texorpdfstring{\Bs{}}{B(s)}, and \texorpdfstring{\PbgLp}{Lambda(b)} hadrons}

The \Bu{}, \Bd{}, \Bs{}, and \PbgLp{} hadrons are reconstructed in the decays \BpJpsiKp{}, \BdJpsiKshort{}, \BdJpsiKstar{}, \BsJpsiPiPi{}, \BsJpsiPhi{}, and \LambJpsiLam. The \Kstar, \PKzS, \Pgf, and \Lamb{} candidates are reconstructed from pairs of oppositely charged tracks that are consistent with originating from a
common vertex.  Because of the lack of charged particle identification, the labelling of tracks as pions, kaons, and protons simply means the mass that is assigned to the track.
The mass assignments for the \PKzS{} and \Pgf{} decay products are unambiguous (either both pions or both kaons).
For the kinematic region considered in this analysis, simulations show that the proton always corresponds to
the track with the larger momentum (leading track) from the \Lamb{} decay.
The \Kstar{} candidates are constructed from a pair of tracks with kaon and pion mass assignments.

Since two \Kstar{} candidates can be formed with a single pair of tracks, we select the combination for which the mass of the \Kstar{} candidate is closest to the world-average value~\cite{PDG_lifetimes16}. This selects the correct combination 88\% of the time.

{\tolerance=1200
All tracks must have a transverse momentum greater than 0.5\GeV. The decay vertices of the \PKzS{} and \Lamb{} particles are required to have a transverse decay length
larger than 15$\sigma$ and their two decay products must each have a transverse impact parameter of at least 2$\sigma$, where the distances are with respect to the beamspot and $\sigma$ is the calculated uncertainty in the relevant quantity.
The intermediate candidate states \Kstar, \PKzS, \Pgf, and \Lamb{} are selected if they lie within the following mass regions that correspond to 1--2 times the experimental resolution or natural width around the nominal mass:
$0.7960<M(\PKp \Pgpm)<0.9880$\GeV, $0.4876<M(\Pgpp \Pgpm)<0.5076$\GeV,
$1.0095 < M(\PKp \PKm) < 1.0295$\GeV, and $1.1096<M(\Pp\Pgpm)<1.1216$\GeV. The accepted mass region of the $\Pgpp \Pgpm$ system in \BsJpsiPiPi{} decay is $0.9240 < M(\Pgpp \Pgpm) < 1.0204$\GeV.
The \PKzS{} contamination in the \Lamb{} sample is removed by discarding candidates in which the leading particle in the
\Lamb{} decay is assigned the pion mass and the resulting $\Pgpp\Pgpm$ invariant mass is in the range $0.4876<M(\Pgpp\Pgpm)<0.5076$\GeV.
Conversely, the \Lamb{} contamination is removed from the \PKzS{} sample by discarding candidates in the $\Pp\Pgpm$ mass region
$1.1096<M(\Pp\Pgpm)<1.1216$\GeV, when the proton mass is assigned to the leading pion from the \PKzS{} decay.
The \pt{} of the \PKp{} candidate track from the \Bu{} decay must be larger than 1\GeV.
The \pt{} of the $\Pgpp \Pgpm$ system in \BsJpsiPiPi{} decays and the \Kstar{} candidates in \BdJpsiKstar{} decays must be greater than 3.5\GeV, with the leading (trailing) charged hadrons in these decays required to have a \pt greater than 2.5 (1.5)\GeV. The \pt{} of the \cPqb{} hadrons must be at least 13\GeV, except for the \BsJpsiPhi{} decay where no requirement is imposed.
The \pt{} of the leading track from the \PKzS{} and \Lamb{} decays must be larger than 1.8\GeV. The minimum \pt{} for the kaons forming a \Pgf{} candidate is 0.7\GeV.
\par}

The \cPqb{} hadron vertex $\chi^2$ probability is required to be greater than 0.1\% in the \BsJpsiPhi{} channel only. The lifetime measurement is limited to events in which the \cPqb{} hadron \ct is greater than 0.02\cm to avoid resolution and reconstruction effects present in the low-\ct region.  No attempt is made to select a single b hadron candidate in the relatively rare $(<1\%)$ events in which more than one b hadron candidate is found.

\subsection{Reconstruction of \texorpdfstring{\BcJpsiPip}{B(c)+ to J/psi pi+}}
\label{sec:BcBprec}

The \PBc lifetime is measured using the method developed by the LHCb Collaboration~\cite{BcLifetimeLHCBhad} in which the measured difference in total widths between the \PBc and \PBp{} mesons is used in combination with the precisely known \PBp{} lifetime to obtain the \PBc lifetime. This method does not require modelling the background \ct distribution, avoiding a source of systematic uncertainty.
The same reconstruction algorithm and selection criteria are used for both decays, \BcJpsiPip and \BJpsiKp.  As a result, the dependence of the efficiencies on the proper decay time is similar.

The charged hadron tracks are required to have at least 2 pixel hits, at least 6 tracker hits (strips and pixels together), a track fit $\chi^2$ less than 3 times the number of degrees of freedom, and $|\eta|< 2.4$.
The dimuon invariant mass is required to lie in the range $\pm$3$\sigma$ from the nominal \cPJgy{} meson mass, where $\sigma$ is the average resolution for the \cPJgy{} signal, which depends on the \cPJgy{} pseudorapidity and ranges from 35 to 50\MeV.
The \pt of the charged hadron tracks and the b hadrons are required to be greater than
3.3 and 10\GeV, respectively.
The b hadrons must have a rapidity of $|y|<2.2$, a vertex $\chi^2$
probability greater than 5\%, a dimuon vertex $\chi^2$ probability greater than 1\%, and
$\cos\theta > 0.98$, where ${\cos\theta = \vec{L}_{xy}\cdot \vec{p}_{\mathrm{T},\PB} / (|L_{xy}|\cdot|p_{\mathrm{T},\PB}|)}$ and $\vec{L}_{xy}$ and $\vec{p}_{\mathrm{T},\PB}$ refer to the
transverse decay length and momentum of the \PBp{} or \PBc mesons. The lifetime measurement is
limited to events in which the b hadron has $\ct > 0.01\cm$, which ensures that the ratio of the  \PBc to \PBp{} meson efficiencies is constant versus \ct.
The analysis of the \PBc lifetime is described in Section~\ref{Sec:Bc}.

\section{Measurement of the \texorpdfstring{\Bd{}}{B0}, \texorpdfstring{\Bs{}}{B0(s)}, and \texorpdfstring{\PbgLp{}}{Lambda(b)} lifetimes} \label{sec:modeling}

For each decay channel, we perform a simultaneous fit to three input variables, the b hadron mass, \ct, and  \ct uncertainty ($\sigma_{ct}$). For the \Bu, \Bd, and \PbgLp{} hadrons, an unbinned maximum-likelihood fit is performed with a probability density function (PDF) given by:
\ifthenelse{\boolean{cms@external}}{
\begin{multline}
\mathrm{PDF} =  f_s\; M_{s}(M)\; T_{s}(\ct)\; E_{s}(\sigma_{\ct})\; \varepsilon(ct) \\
+ (1-f_s)\; M_{b}(M)\; T_{b}(\ct)\; E_{b}(\sigma_{\ct}),
\end{multline}
}{
\begin{equation}
\mathrm{PDF} =  f_s\; M_{s}(M)\; T_{s}(\ct)\; E_{s}(\sigma_{\ct})\; \varepsilon(ct) + (1-f_s)\; M_{b}(M)\; T_{b}(\ct)\; E_{b}(\sigma_{\ct}),
\end{equation}
}
where $f_s$ is the fraction of signal events, and $M_s$ ($M_b$), $T_s$ ($T_b$), and $E_s$ ($E_b$) are the functions describing the signal (background) distributions of the b hadron mass, \ct, and $\sigma_{\ct}$, respectively, while $\varepsilon$ is the efficiency function. These functions are derived below. For the \Bs{} modes, we use an extended maximum-likelihood fit in order to correctly incorporate background sources whose yields are obtained from the fit.

\subsection{Reconstruction and selection efficiency}

The reconstruction and selection efficiency $\varepsilon$ for each decay mode is determined as a function of \ct by using fully simulated MC samples. This efficiency is defined as the generated \ct{} distribution of the selected events after reconstruction and selection divided by the \ct distribution obtained from an exponential decay with the lifetime set to the value used to generate the events. The efficiency for the \BsJpsiPhi{} channel is defined as the generated \ct  distribution of the selected events after reconstruction divided by the sum of the two exponentials generated with the theoretical \BsJpsiPhi{} decay rate model~\cite{Dighe:1998vk}. In the theoretical model, the values of the physics parameters are set to those used in the simulated sample.

\begin{figure*}[htbp]
\centering
       \includegraphics[width=0.48\textwidth]{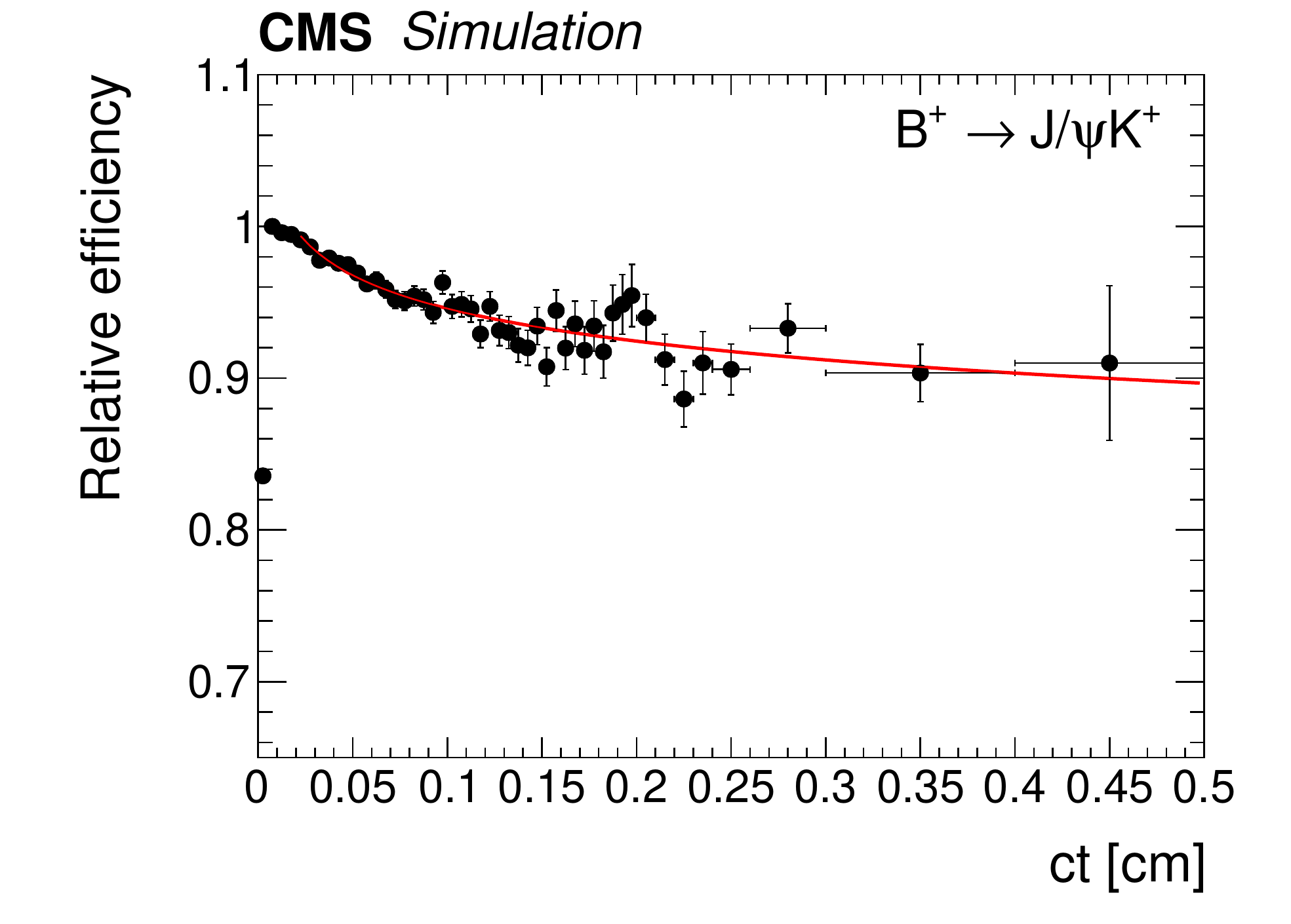}
       \includegraphics[width=0.48\textwidth]{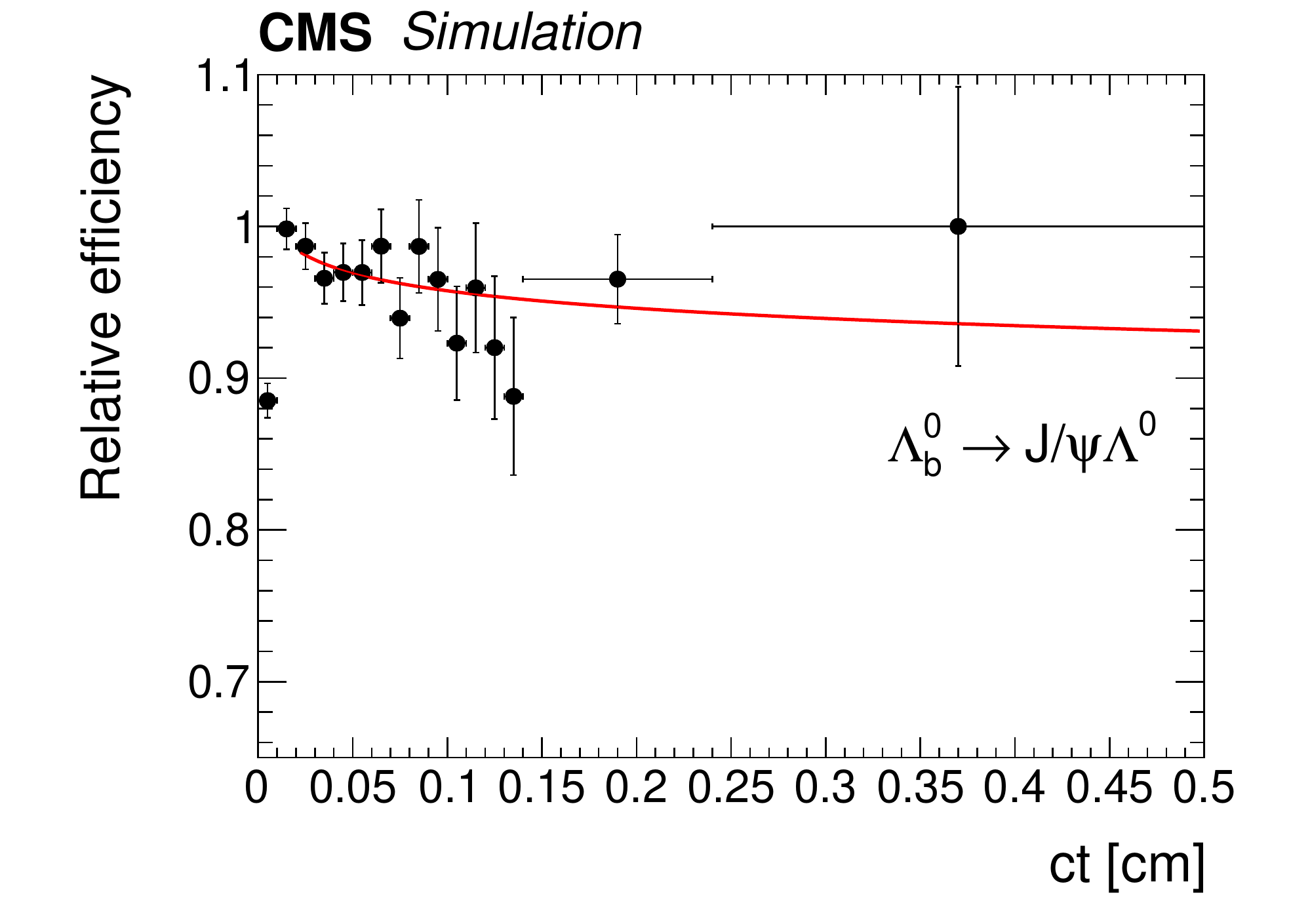}
       \includegraphics[width=0.48\textwidth]{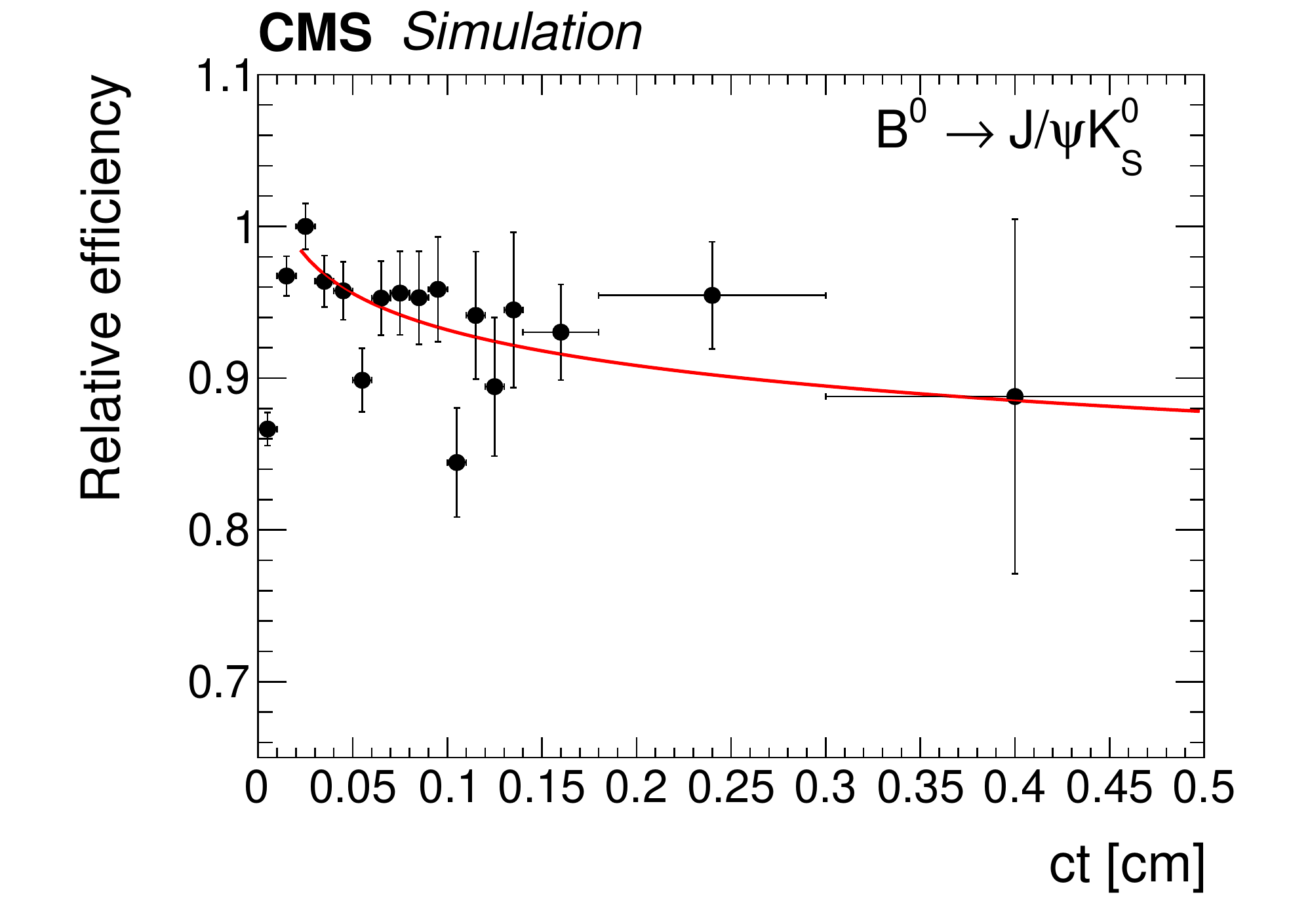}
       \includegraphics[width=0.48\textwidth]{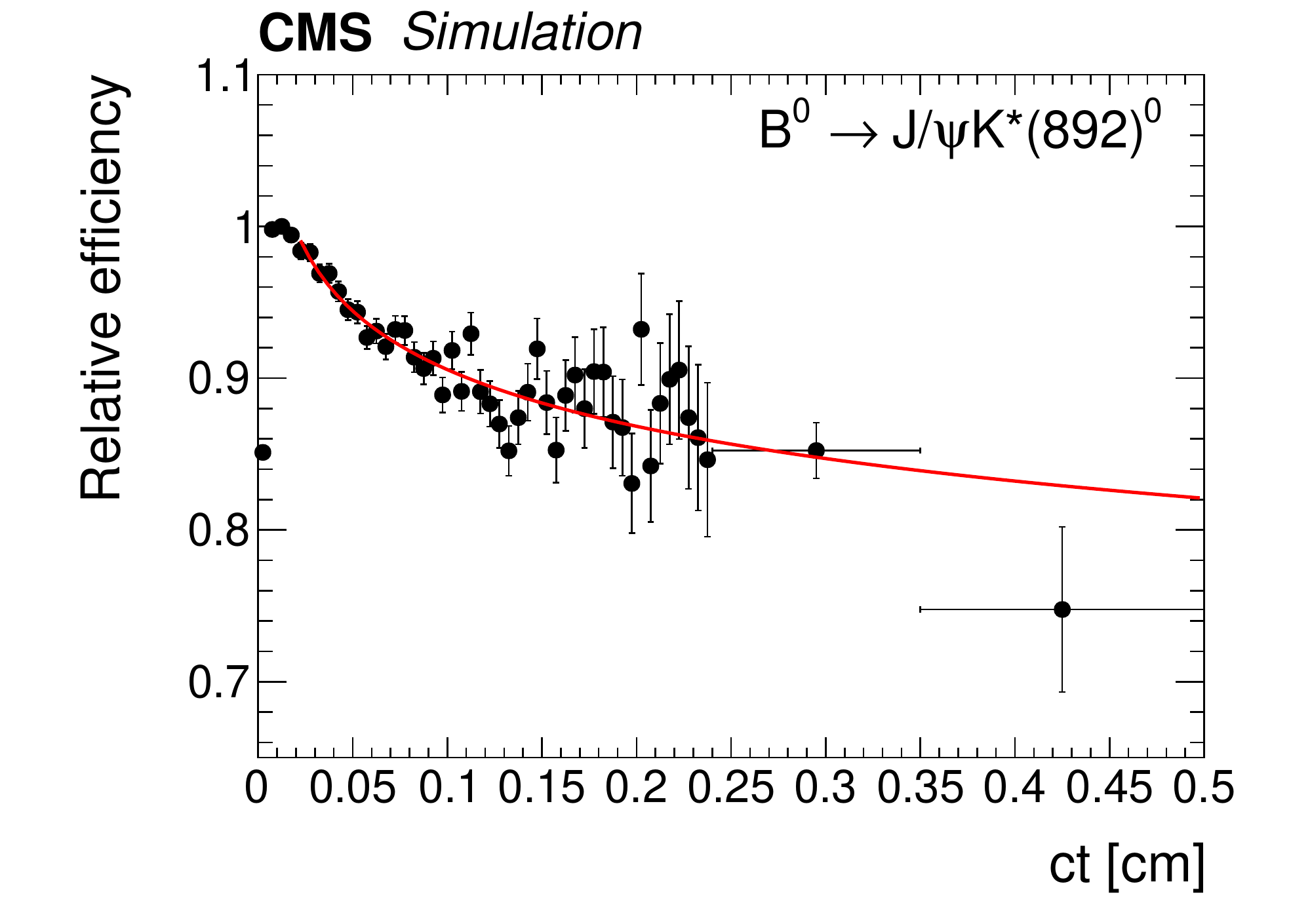}
       \includegraphics[width=0.48\textwidth]{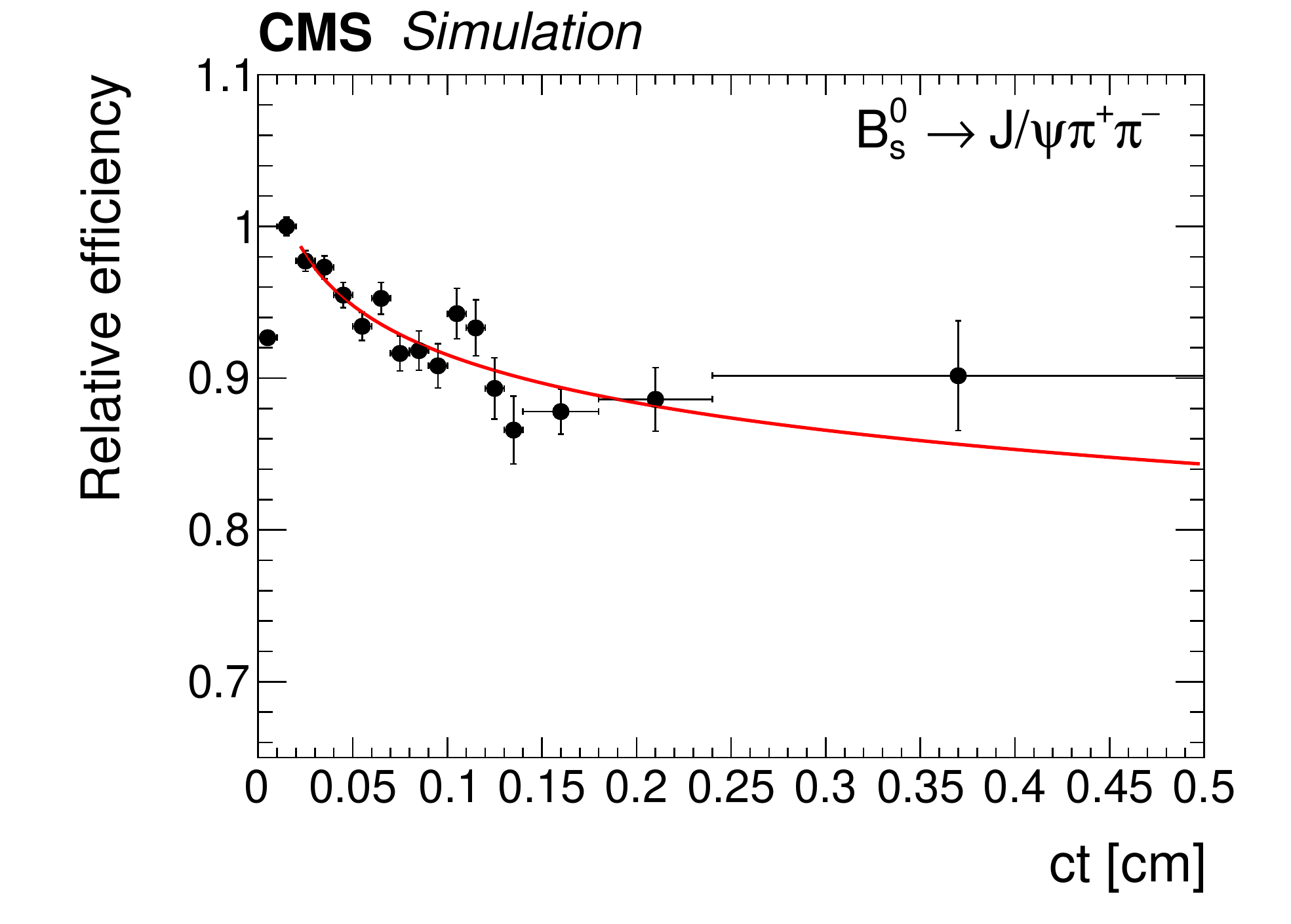}
       \includegraphics[width=0.48\textwidth]{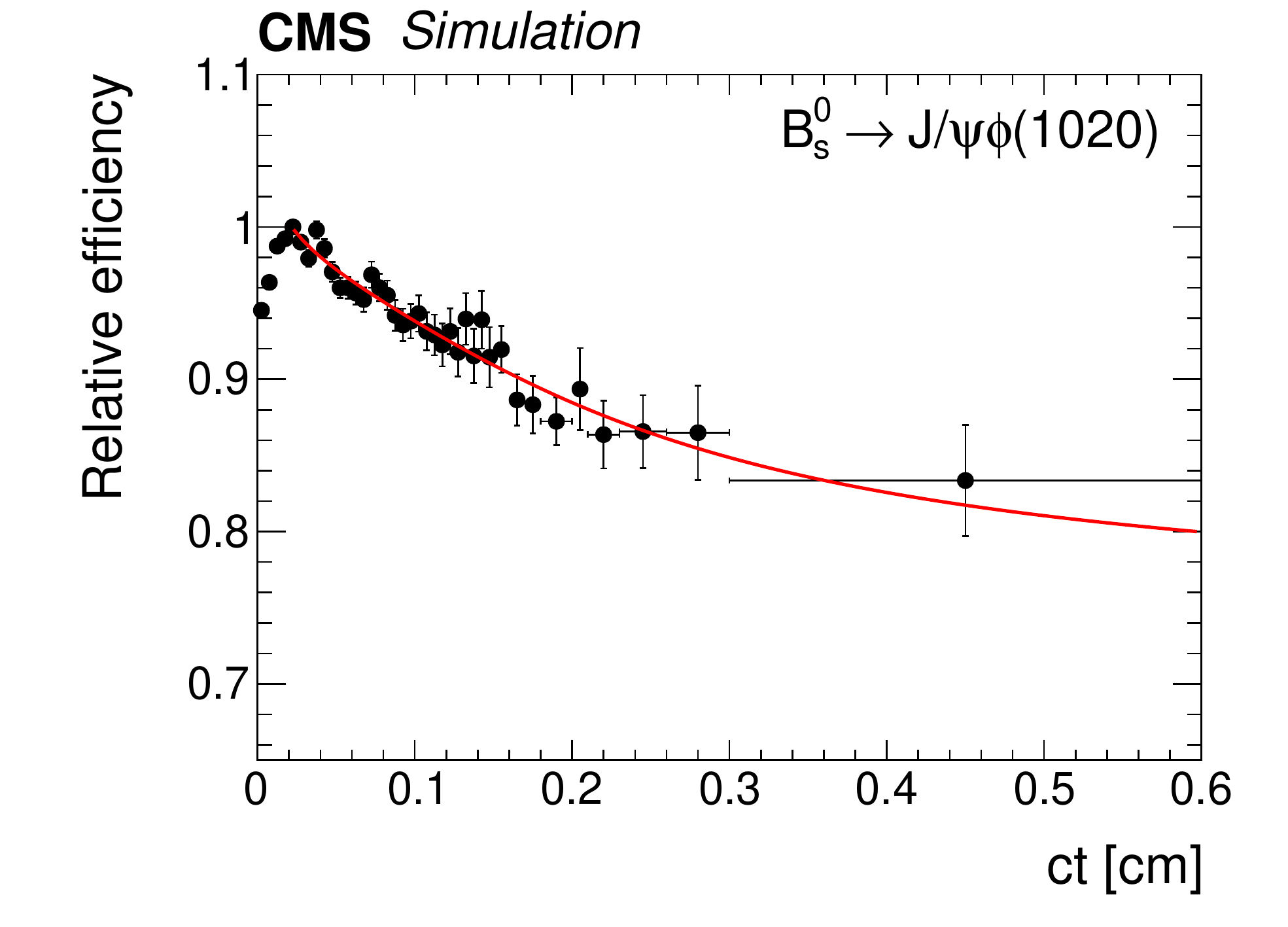}
\caption{ \label{fig:TriggerCorrection} The combined reconstruction and selection efficiency from simulation versus \ct{} with a superimposed fit to an inverse power function for \BpJpsiKp{} (upper left), \LambJpsiLam{} (upper right), \BdJpsiKshort{} (centre left), \BdJpsiKstar{} (centre right), \BsJpsiPiPi{} (lower left), and \BsJpsiPhi{} (lower right). The efficiency scale is arbitrary.}
\end{figure*}

Figure~\ref{fig:TriggerCorrection} shows the efficiency as a function of \ct{} for the various decay modes, with an arbitrary normalization since only the relative efficiency is relevant. The efficiencies display a sharp rise as \ct{} increases from 0 to 0.01\unit{cm}, followed by a slow decrease as \ct{} increases further. The \ct{} efficiency is modelled with an inverse power function.

\subsection{Data modelling}\label{sec:datamodeling}

Depending on the decay channel, the invariant mass distribution for the signal $M_s$ is modelled with one or two Gaussian functions, and a linear polynomial or an exponential function is used to model the combinatorial background $M_b$. For the \BsJpsiPiPi{} decay, three additional terms are added to $M_b$ to include specific sources of background. The \BdJpsiPiPi{} decays are modelled by a Gaussian function, the \BpJpsiKp{} decays by a shape taken from simulation, and the $\PBz_{\mathrm{(d,s)}} \!\to\! \Jpsi \mathrm{h}^{+}_{1} \mathrm{h}^{-}_{2}$ decays, where $\mathrm{h}^{+}_{1}$ and $\mathrm{h}^{-}_{2}$ are charged hadron tracks that are not both pions, by a Gaussian function.

The signal \ct{} distribution $T_s$ is modelled by an exponential function convolved with
the detector resolution and then multiplied by the function describing the reconstruction and selection efficiency. The resolution is described by a Gaussian function with the per-event width taken from the \ct uncertainty distribution.  The backgrounds $T_b$ are described by a superposition of exponential functions convolved with the resolution. The number of exponentials needed to describe the background is determined from data events in the mass sideband regions for each decay mode.

The signal $E_s$ and background $E_b$ $\sigma_{ct}$ distributions are modelled with
a sum of two gamma functions for the \BsJpsiPhi{} channel and two exponential functions convolved with a Gaussian function for the other channels.
The background parameters are obtained from a fit to the mass sideband distributions.
The signal parameters are obtained from a fit to the signal region after subtracting the background contribution using the mass sideband region to estimate the background.
The parameters of the efficiency function and the functions modelling the $\sigma_{\ct}$ distributions are kept constant in the fit. The remaining fit parameters are allowed to vary freely.

For the \BsJpsiPiPi{} mode, the parameters of the mass model for the \BpJpsiKp{} contamination are taken from the simulation, and the yield and lifetime
are determined by the fit. The mass of the \BdJpsiPiPi{} contamination is fixed to the weighted average of the masses measured from our two \Bd{} decay modes,
and the width of the Gaussian function is the same as the width used for the \BsJpsiPiPi{} signal, corrected by a factor of $M_{\Bd}/M_{\Bs}$.
The lifetime of this contamination is fixed to the world-average value, corrected by the same factor as the width, and the yield is a free parameter of the fit.

\subsection{Fit results}\label{sec:fitresults}

\begin{figure*} [htbp]
\centering
       \includegraphics[width=0.48\textwidth]{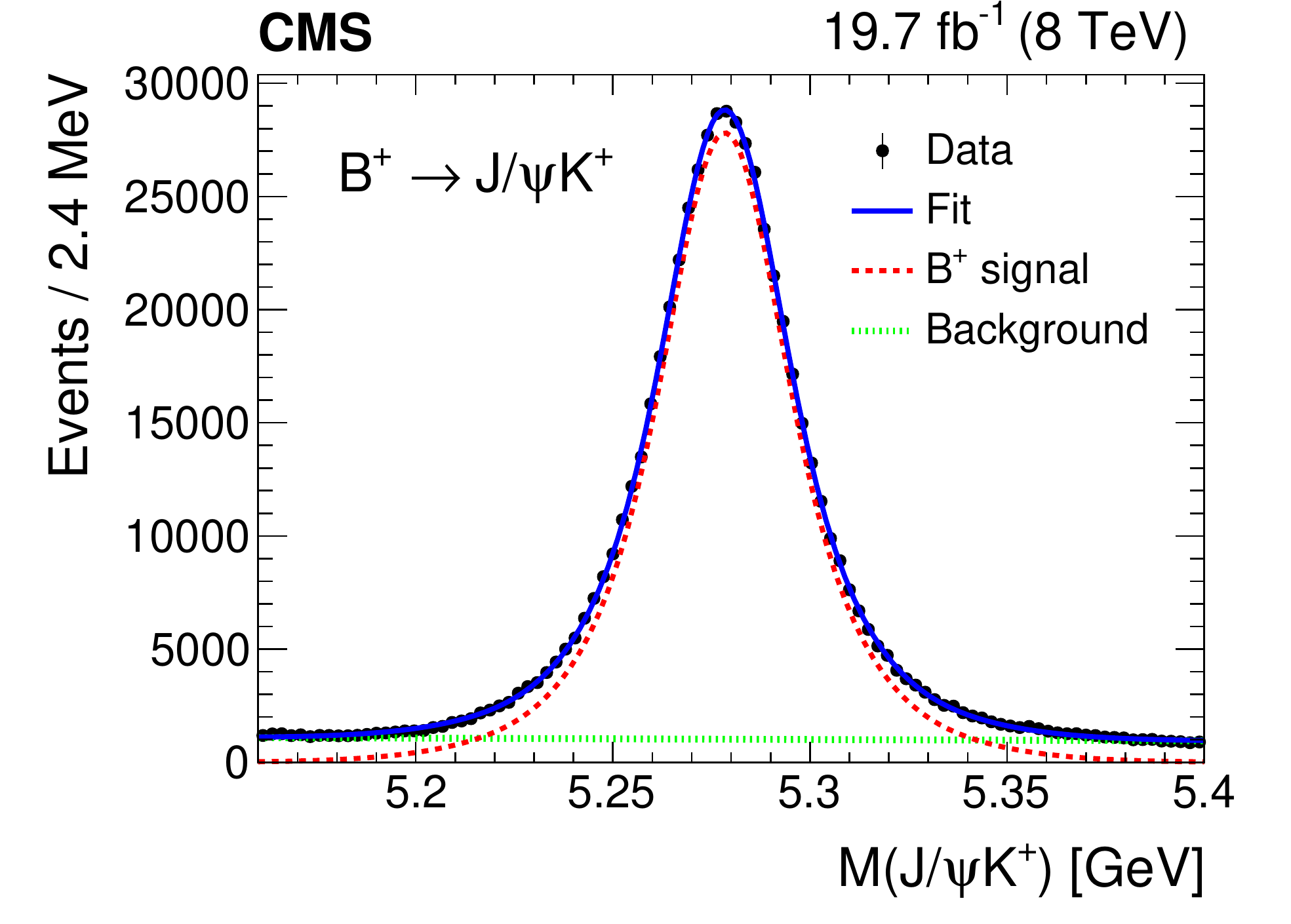}
       \includegraphics[width=0.48\textwidth]{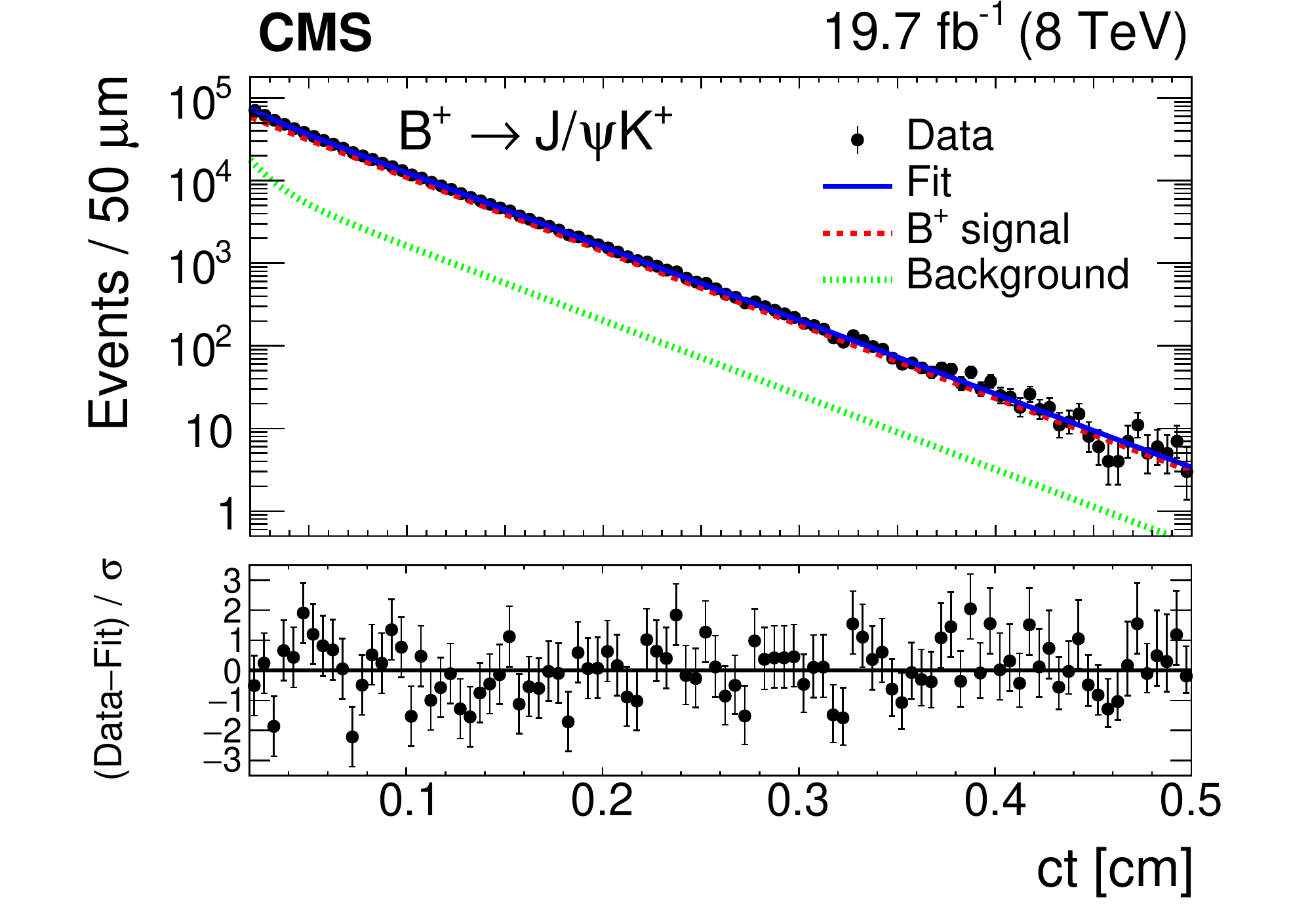}
       \includegraphics[width=0.48\textwidth]{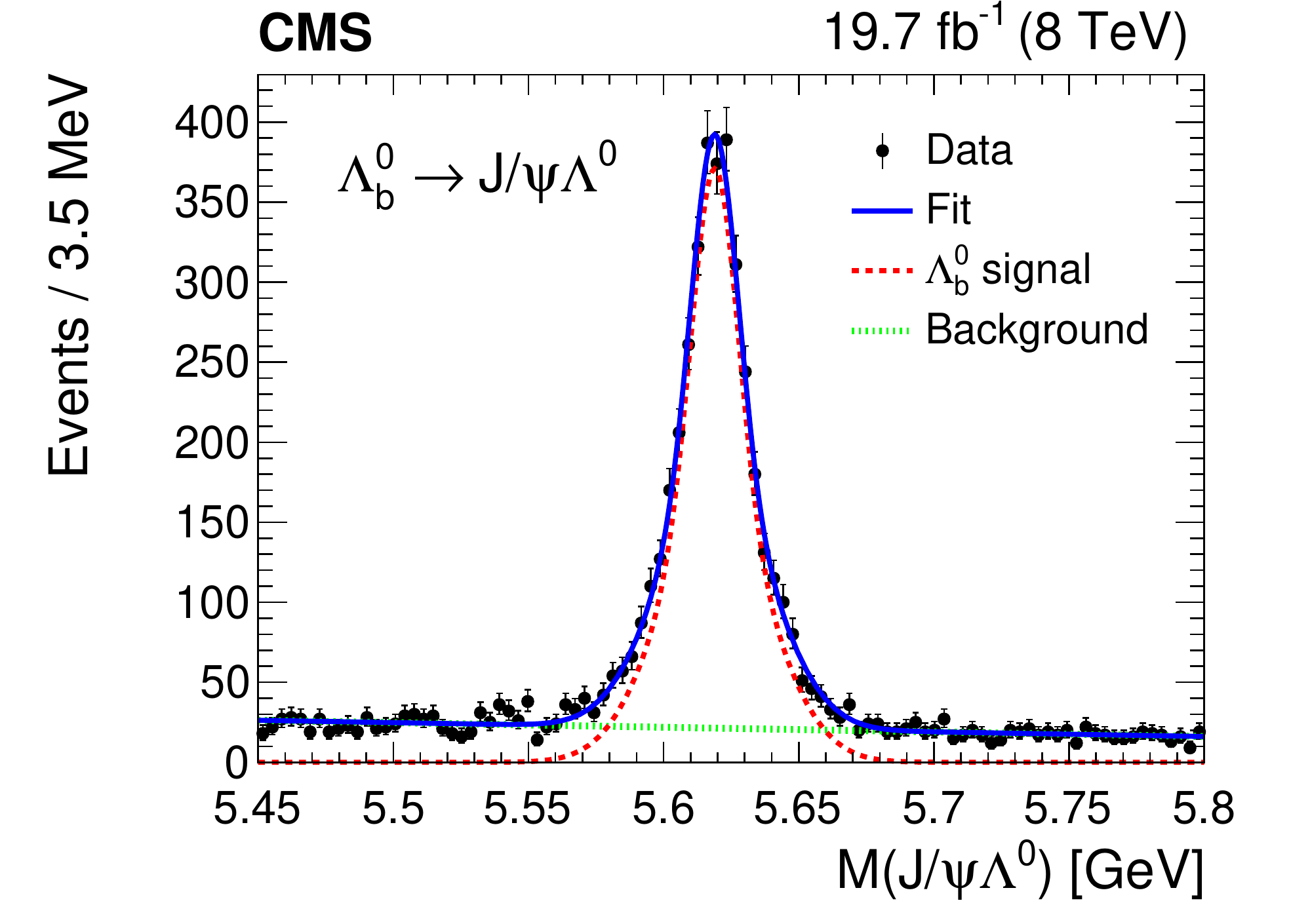}
       \includegraphics[width=0.48\textwidth]{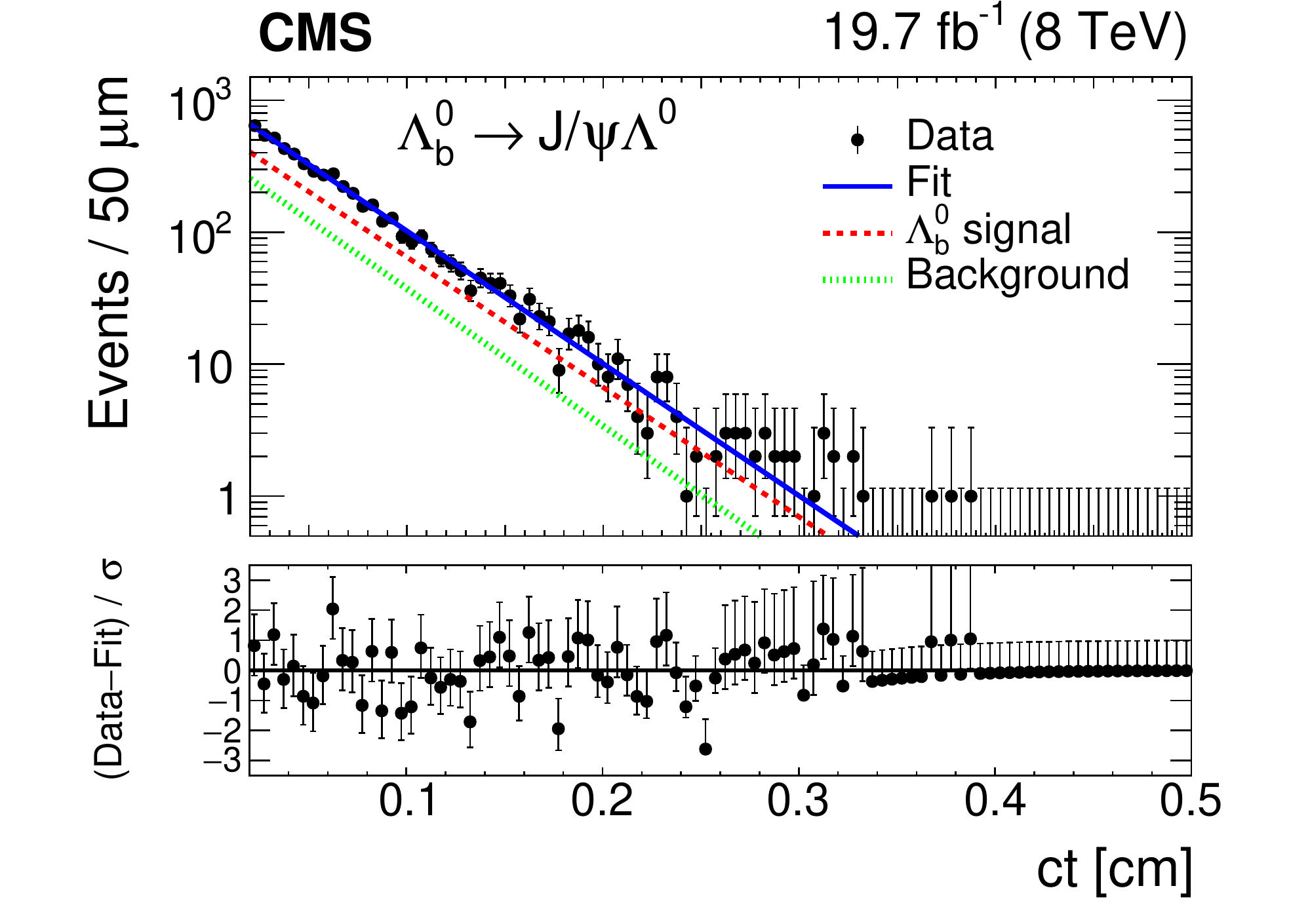}
\caption{ \label{fig:fitBp} Invariant mass (left) and \ct (right) distributions for \Bu{} (upper) and for \PbgLp{} (lower) candidates. The curves are projections of the fit to the data, with the contributions from signal (dashed), background (dotted), and the sum of signal and background (solid) shown. the lower panels of the figures on the right show the difference between the observed data and the fit divided by the data uncertainty.  The vertical bars on the data points represent the statistical uncertainties.
}
\end{figure*}

\begin{figure*} [htbp]
\centering
       \includegraphics[width=0.48\textwidth]{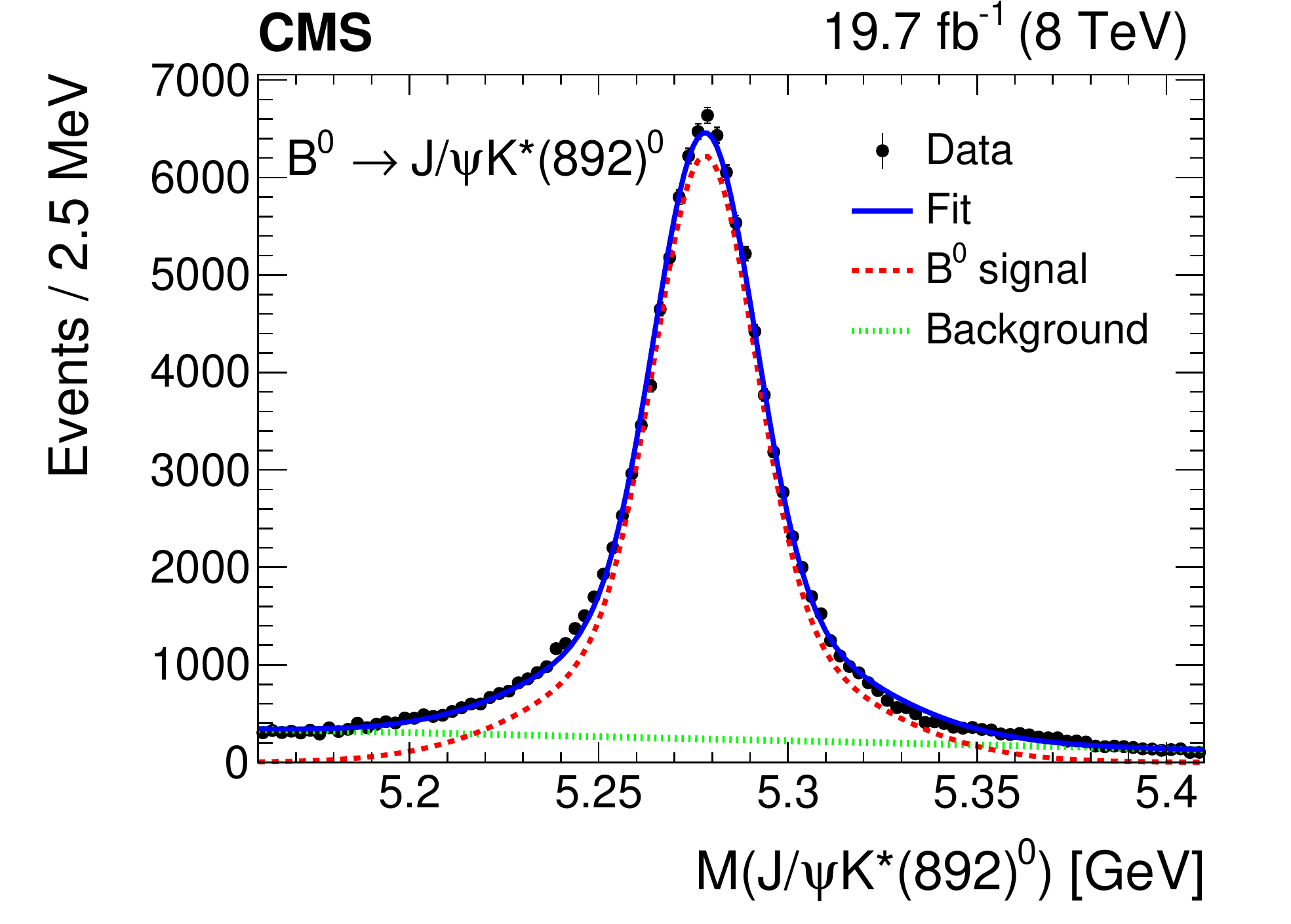}
       \includegraphics[width=0.48\textwidth]{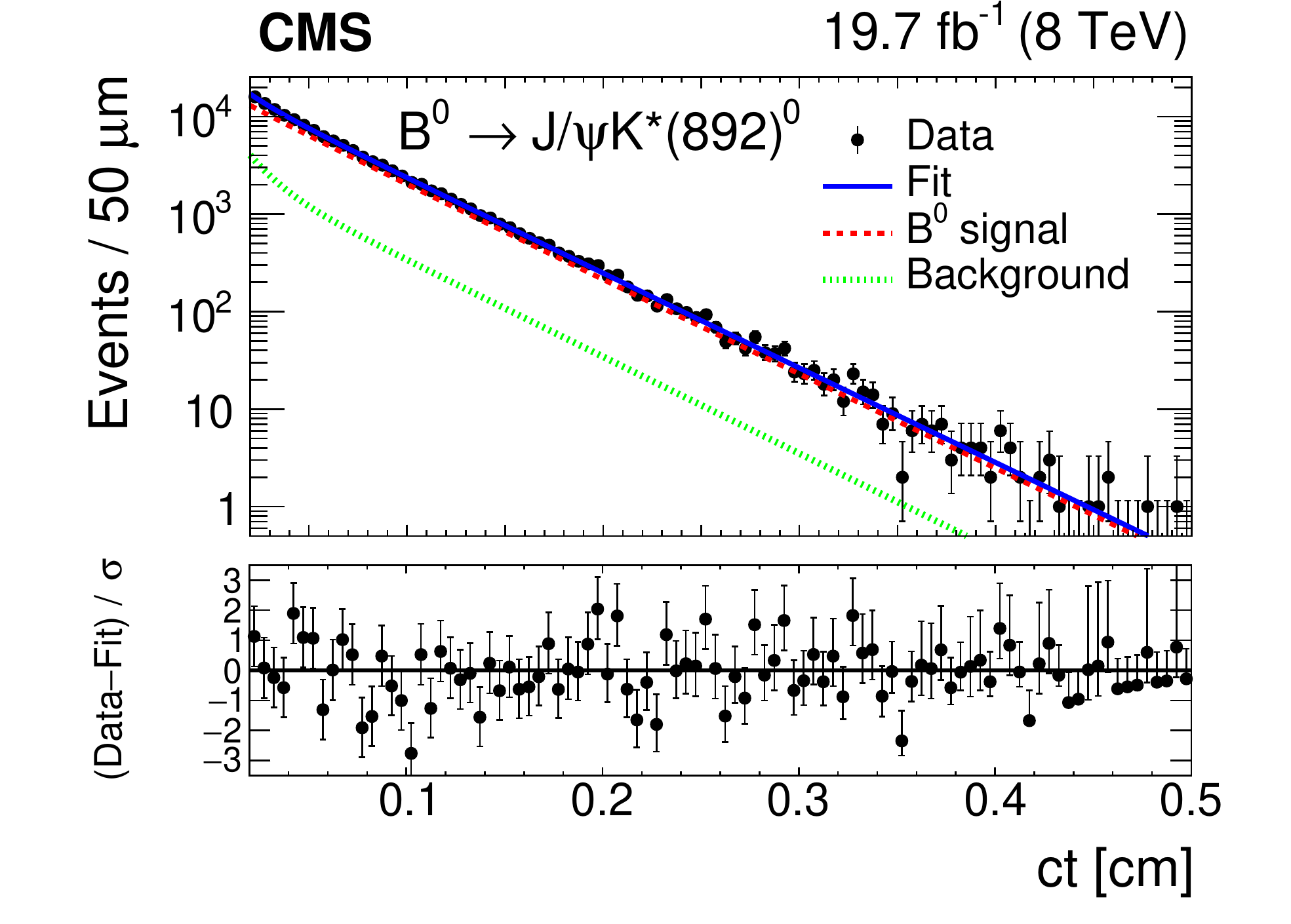}
        \includegraphics[width=0.48\textwidth]{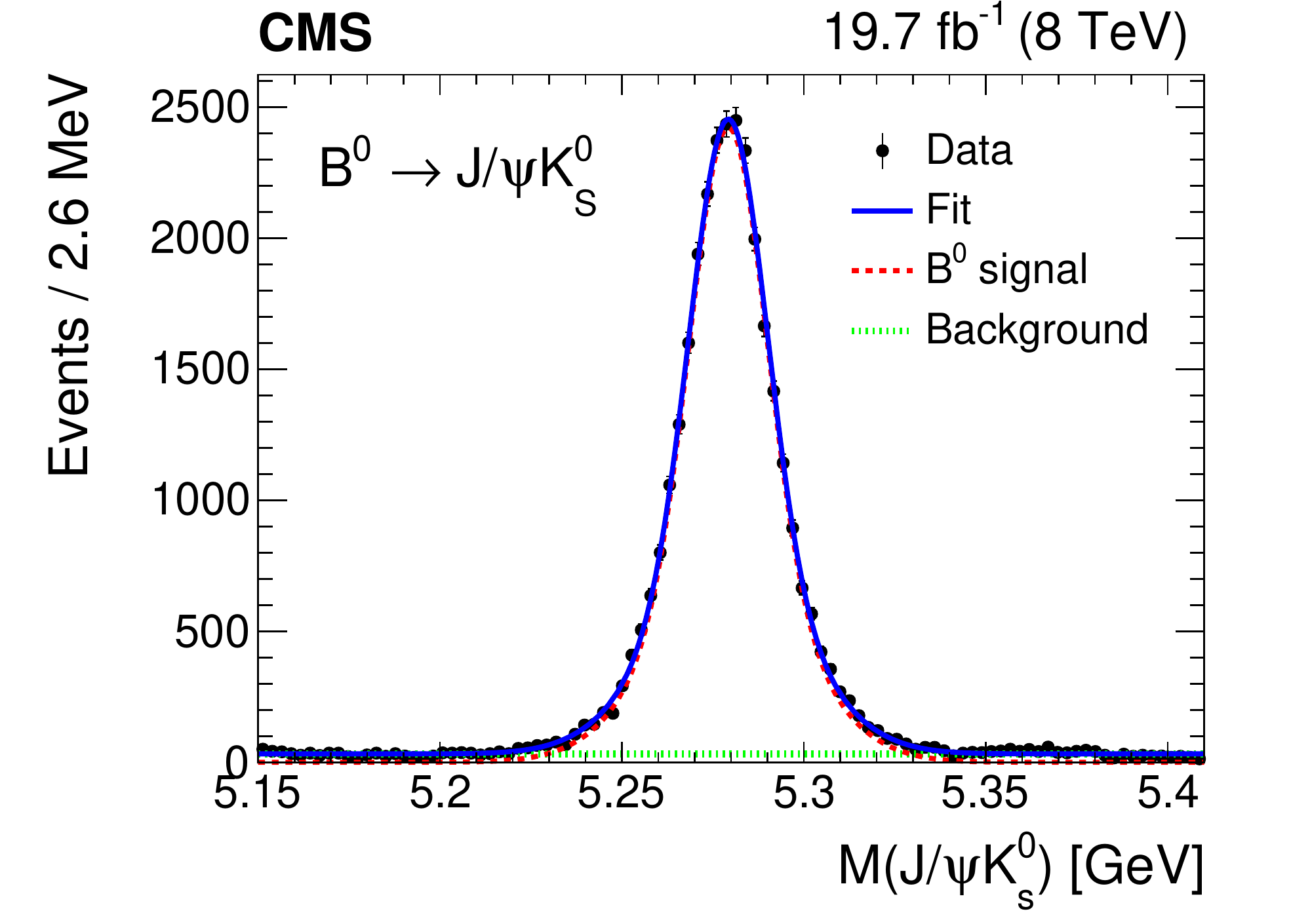}
       \includegraphics[width=0.48\textwidth]{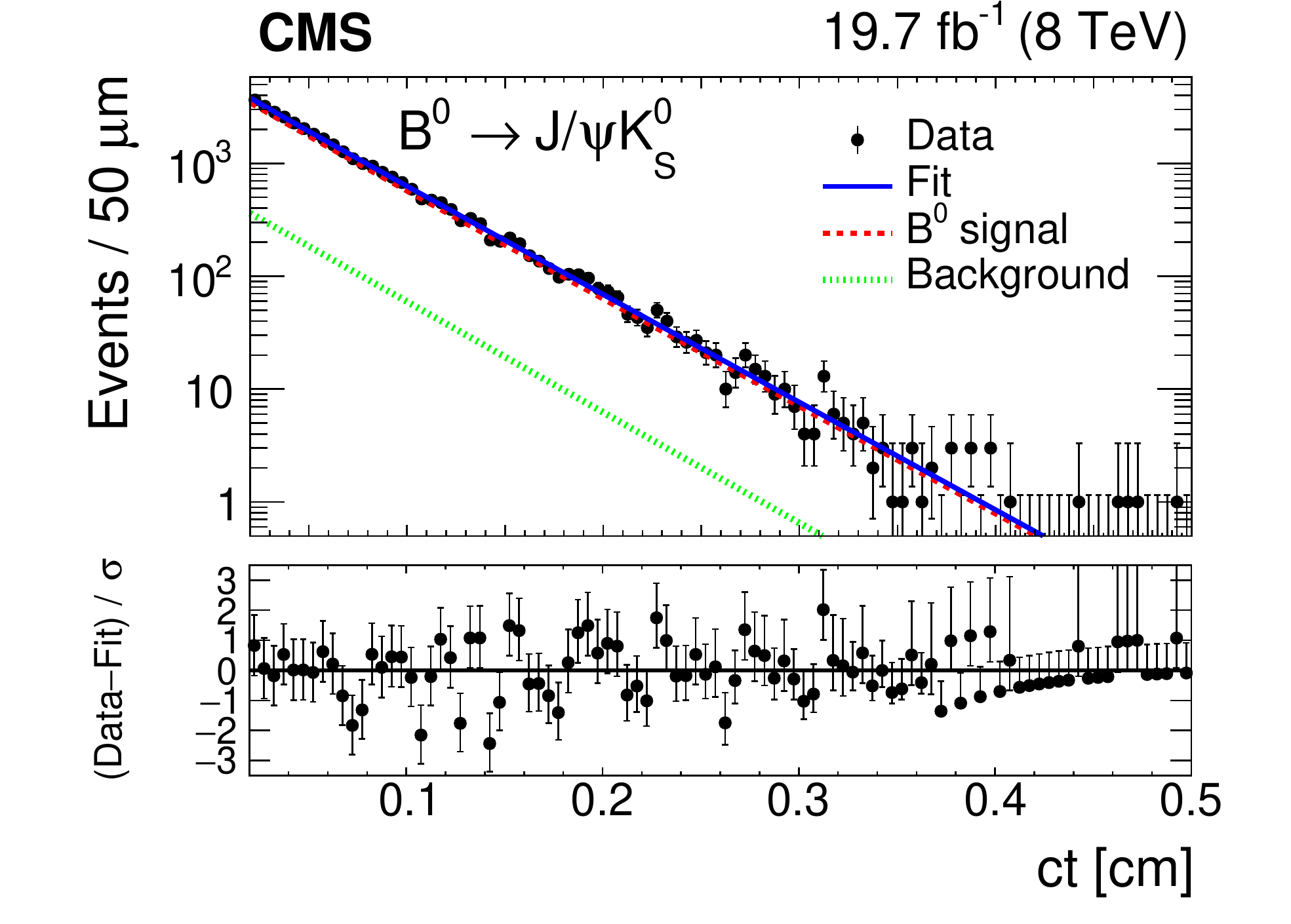}
\caption{ \label{fig:fitBdKstar} Invariant mass (left) and \ct (right) distributions for \Bd{} candidates reconstructed from \JpsiKstar{} (upper) and \JpsiKshort{} (lower) decays. The curves are projections of the fit to the data, with the contributions from signal (dashed), background (dotted), and the sum of signal and background (solid) shown. the lower panels of the figures on the right show the difference between the observed data and the fit divided by the data uncertainty.  The vertical bars on the data points represent the statistical uncertainties.
}
\end{figure*}

\begin{figure*} [htbp]
\centering
       \includegraphics[width=0.48\textwidth]{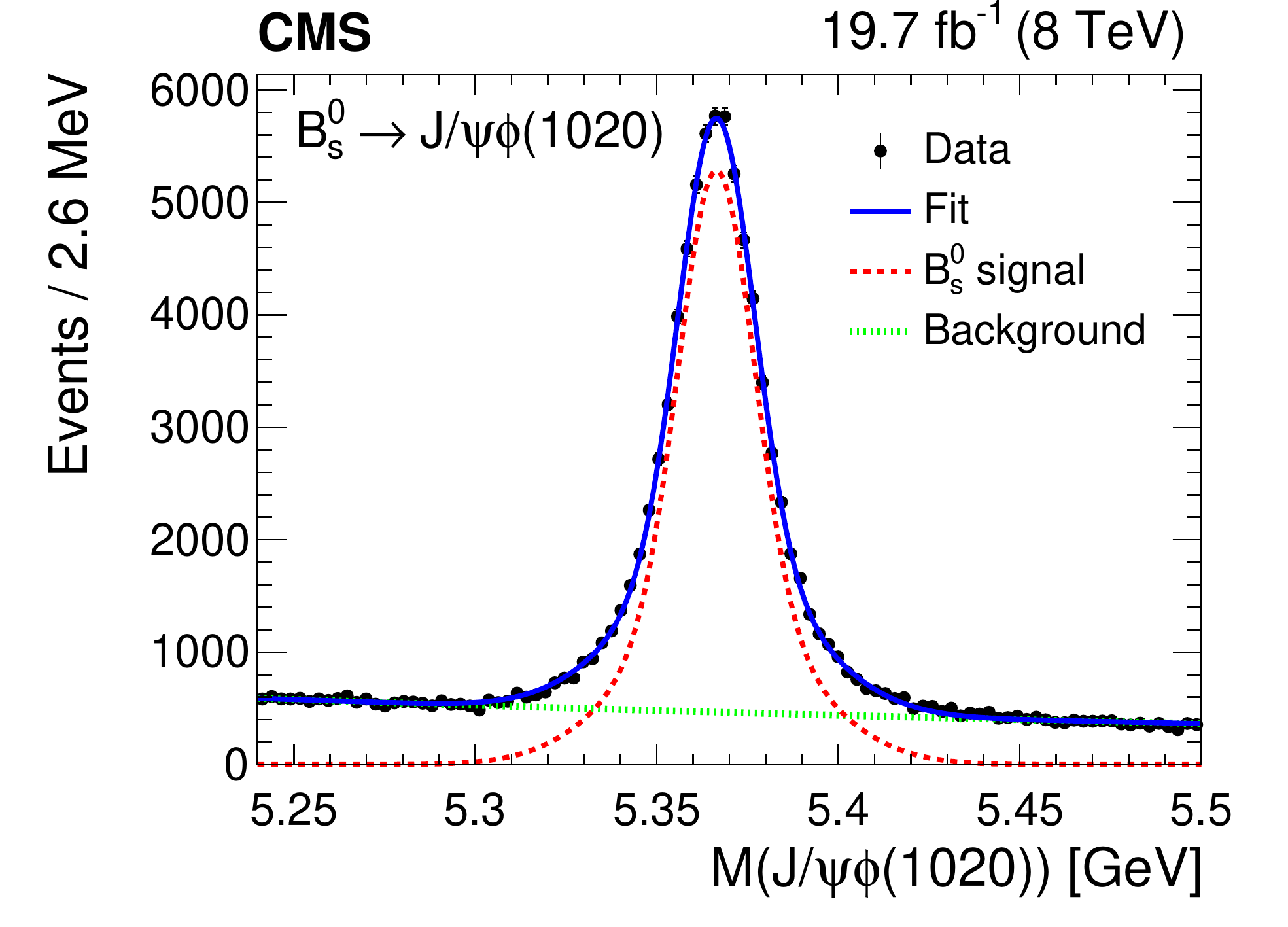}
       \includegraphics[width=0.48\textwidth]{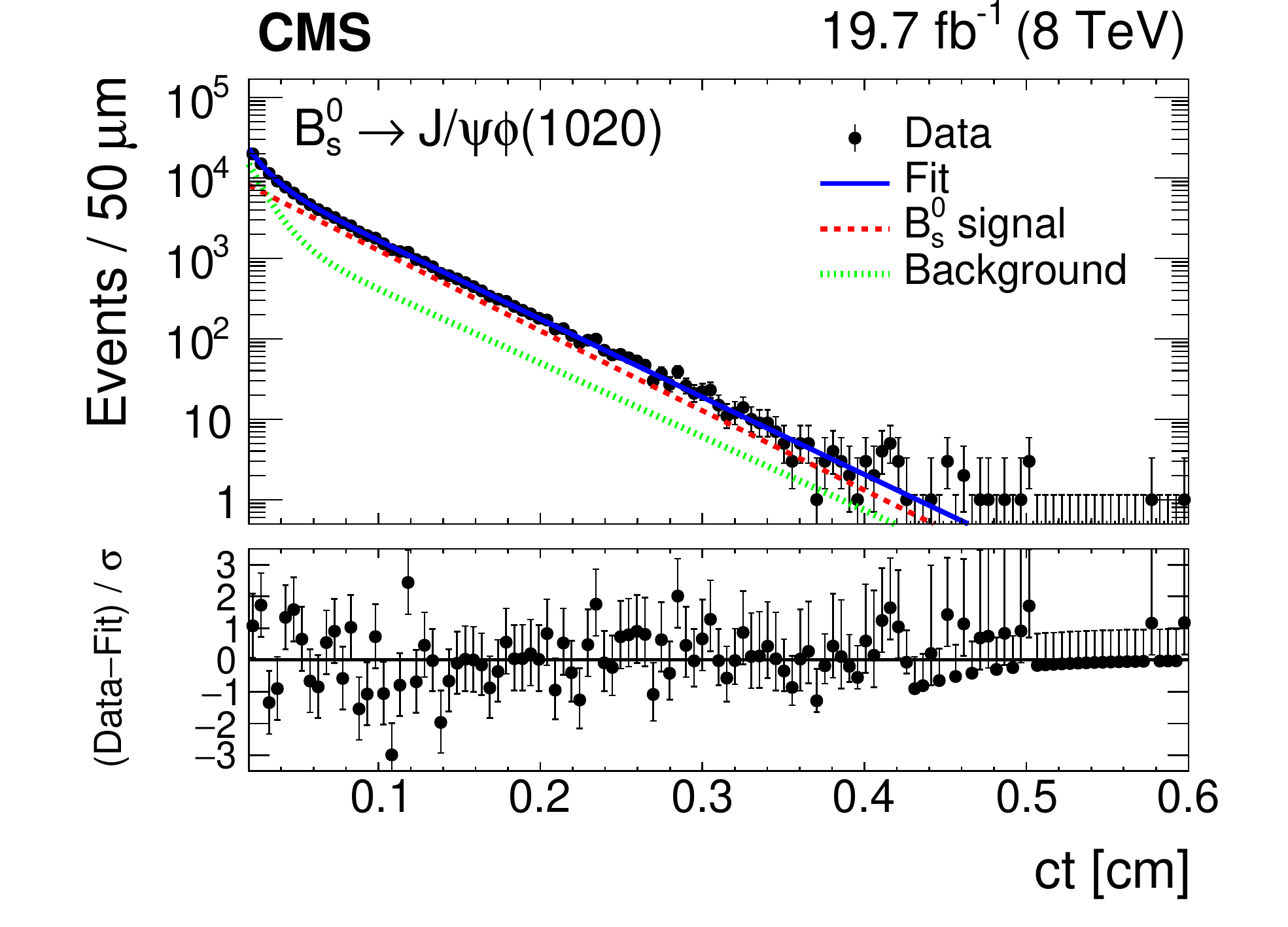}
       \includegraphics[width=0.48\textwidth]{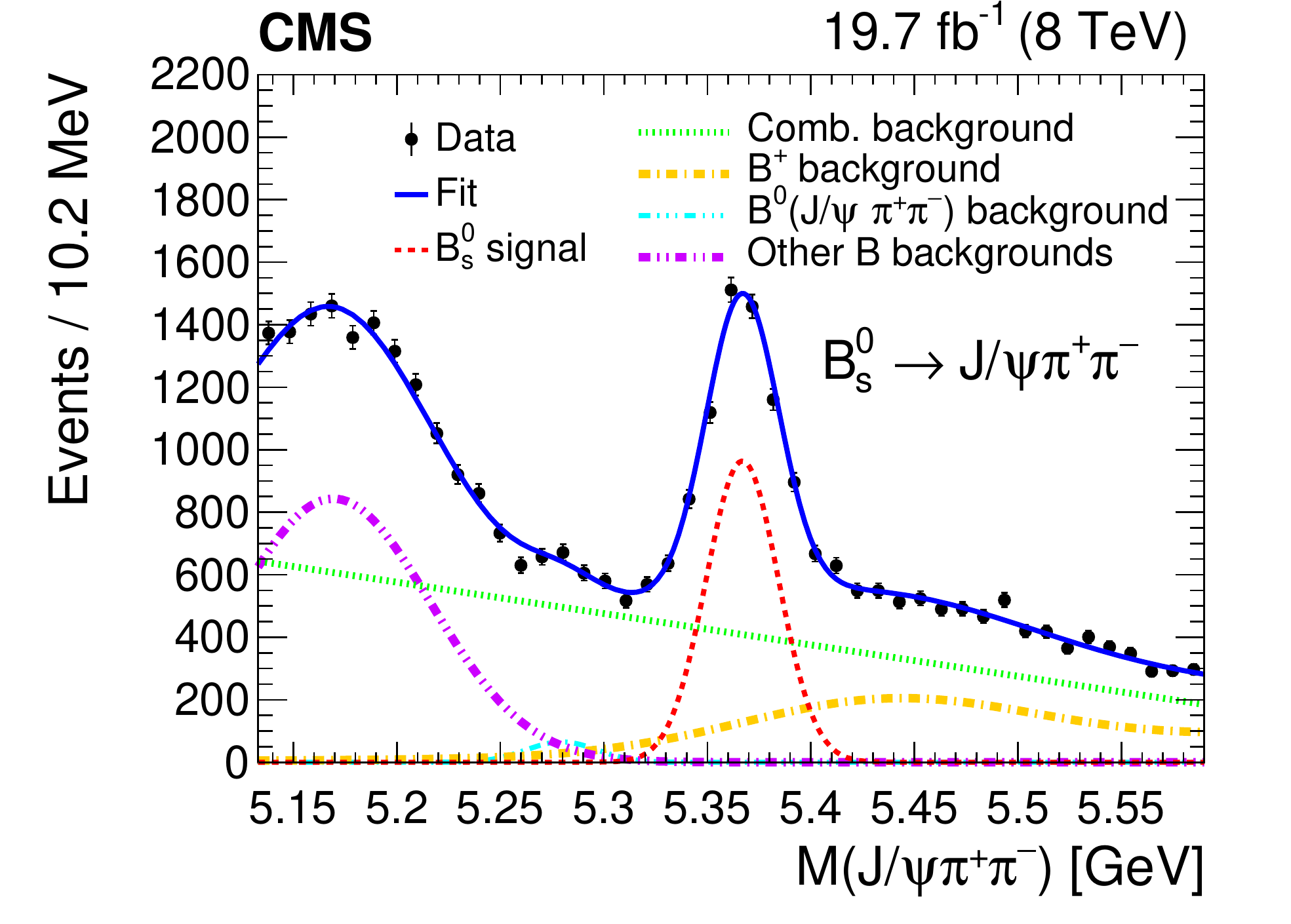}
	   \includegraphics[width=0.48\textwidth]{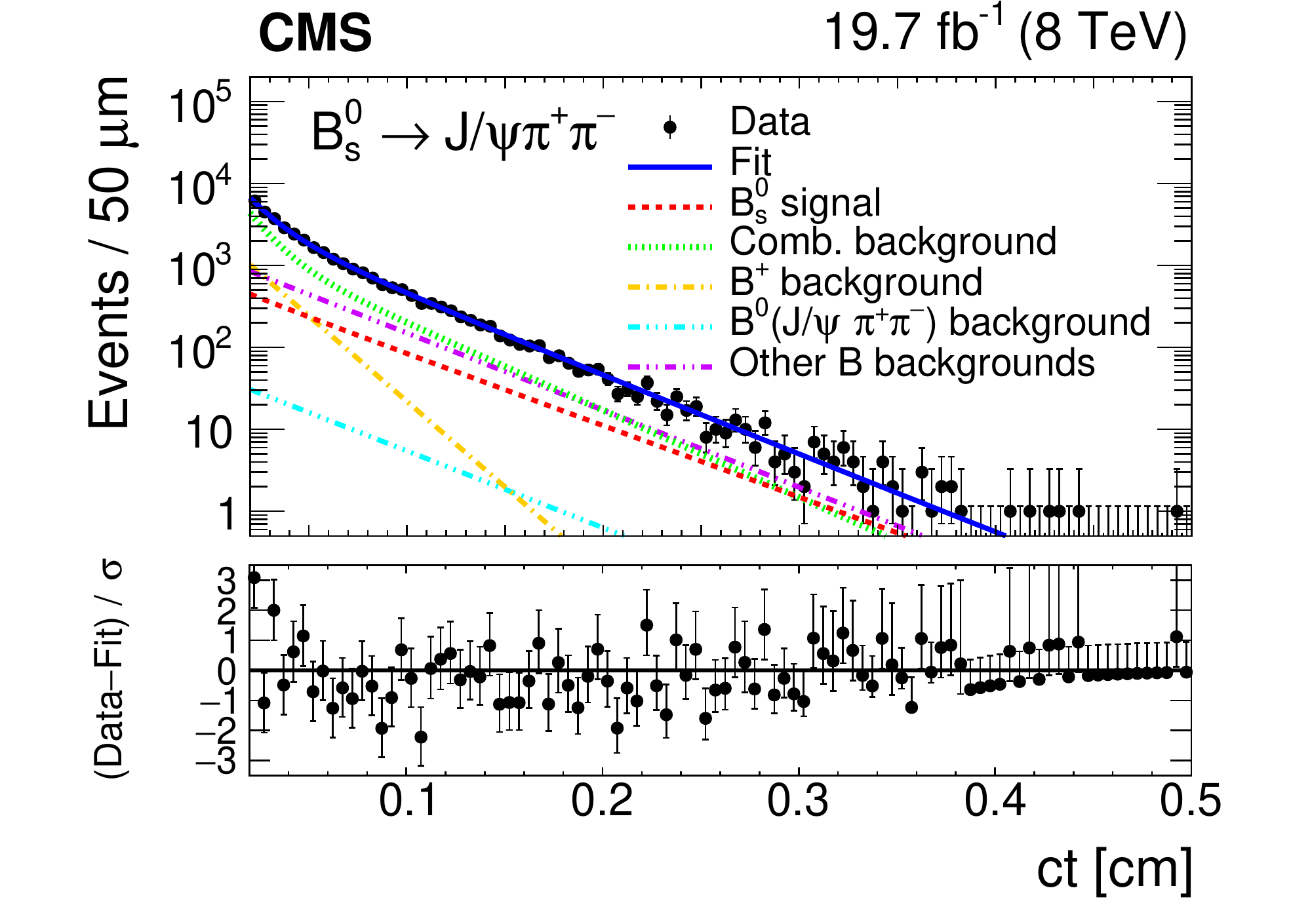}
\caption{ \label{fig:fitBs} Invariant mass (left) and \ct (right) distributions for \Bs{} candidates reconstructed from \JpsiPhi{} (upper) and \JpsiPiPi{} (lower) decays. The curves are projections of the fit to the data, with the full fit function (solid) and signal (dashed) shown for both decays, the total background (dotted) shown for the upper plots, and the combinatorial background (dotted), misidentified \BpJpsiKp{} background (dashed-dotted), \BdJpsiPiPi{} contribution (dashed-dotted-dotted-dotted), and partially reconstructed and other misidentified \PB{} backgrounds (dashed-dotted-dotted) shown for the lower plots. the lower panels of the figures on the right show the difference between the observed data and the fit divided by the data uncertainty. The vertical bars on the data points represent the statistical uncertainties.}
\end{figure*}

The invariant mass and \ct distributions obtained from data are shown with the fit results superimposed in Figs.~\ref{fig:fitBp}--\ref{fig:fitBs}. The \ct distributions are fitted in the range 0.02--0.50\cm for all modes except the \BsJpsiPhi{} channel, where the upper limit is increased to 0.60\cm. The average lifetimes times the speed of light obtained from the fits are: $c\tau_{\Bu} = 490.9 \pm 0.8\micron$,  $c\tau_{\PBz \to \cPJgy \PKst{}^{0}} =  453.0 \pm 1.6\micron$,  $c\tau_{\PBz \to \cPJgy \PKzS}  =  457.8 \pm 2.7\micron$,  $c\tau_{\PBzs \to \cPJgy \Pgpp \Pgpm} = 502.7 \pm 10.2\micron$, $c\tau_{\PBzs \to \cPJgy \Pgf} =  445.2 \pm 2.0\micron$, and $c\tau_{\PbgLp} = 442.9 \pm 8.2\micron$, where all uncertainties are statistical only. The \BsJpsiPhi{} value given here is uncorrected for two offsets described in Section~\ref{sec:syst}. There is good agreement between the fitted functions and the data. The probabilities calculated from the $\chi^2$ of the $\ct$ distributions in Figs.~\ref{fig:fitBp}--\ref{fig:fitBs} all exceed 25\%.

\section{Measurement of the \texorpdfstring{\PBc}{B(c)} lifetime}
\label{Sec:Bc}

The decay time distribution for the signal $N_\mathrm{B}(ct)$ can be expressed as the product of an efficiency function $\varepsilon_\mathrm{B}(ct)$ and an exponential decay function $E_\mathrm{B}(ct) =\exp(-ct/c\tau_\mathrm{B})$, convolved with the time resolution function of the detector $r(ct)$.
The ratio of \PBc to \PBp{} events
at a given proper time can be expressed as
\begin{equation}
\label{Eq:RatioComplete}
\frac {N_{\PBc}(ct)}{N_{\PBp}(ct)} \equiv R(ct)  =  \frac{\varepsilon_{\PBc}(ct) [r(ct)\otimes  E_{\PBc}(ct)]}{\varepsilon_{\PBp}(ct) [r(ct)\otimes E_{\PBp}(ct)]}.
\end{equation}
We have verified through studies of simulated pseudo-events that Eq.~(\ref{Eq:RatioComplete}) is not significantly affected by the time resolution, and therefore this equation can be simplified to
\begin{equation}
\label{Eq:RatioSimplified}
R(ct) \approx R_{\varepsilon}(ct) \exp(-\Delta \Gamma t),
\end{equation}
where the small effect from the time resolution is evaluated from MC simulations and is included in $R_{\varepsilon}(ct)$, which denotes the ratio of the \PBc and \PBp{} efficiency functions.
The quantity $\Delta \Gamma$ is defined as
\begin{equation}
\label{Eq:DeltaGamma}
\Delta \Gamma \equiv \Gamma_{\PBc} - \Gamma_{\PBp} = \frac{1}{\tau_{\PBc}} - \frac{1}{\tau_{\PBp}}.
\end{equation}

\begin{figure}[hbtp]
\centering
    \includegraphics[width=0.48\textwidth]{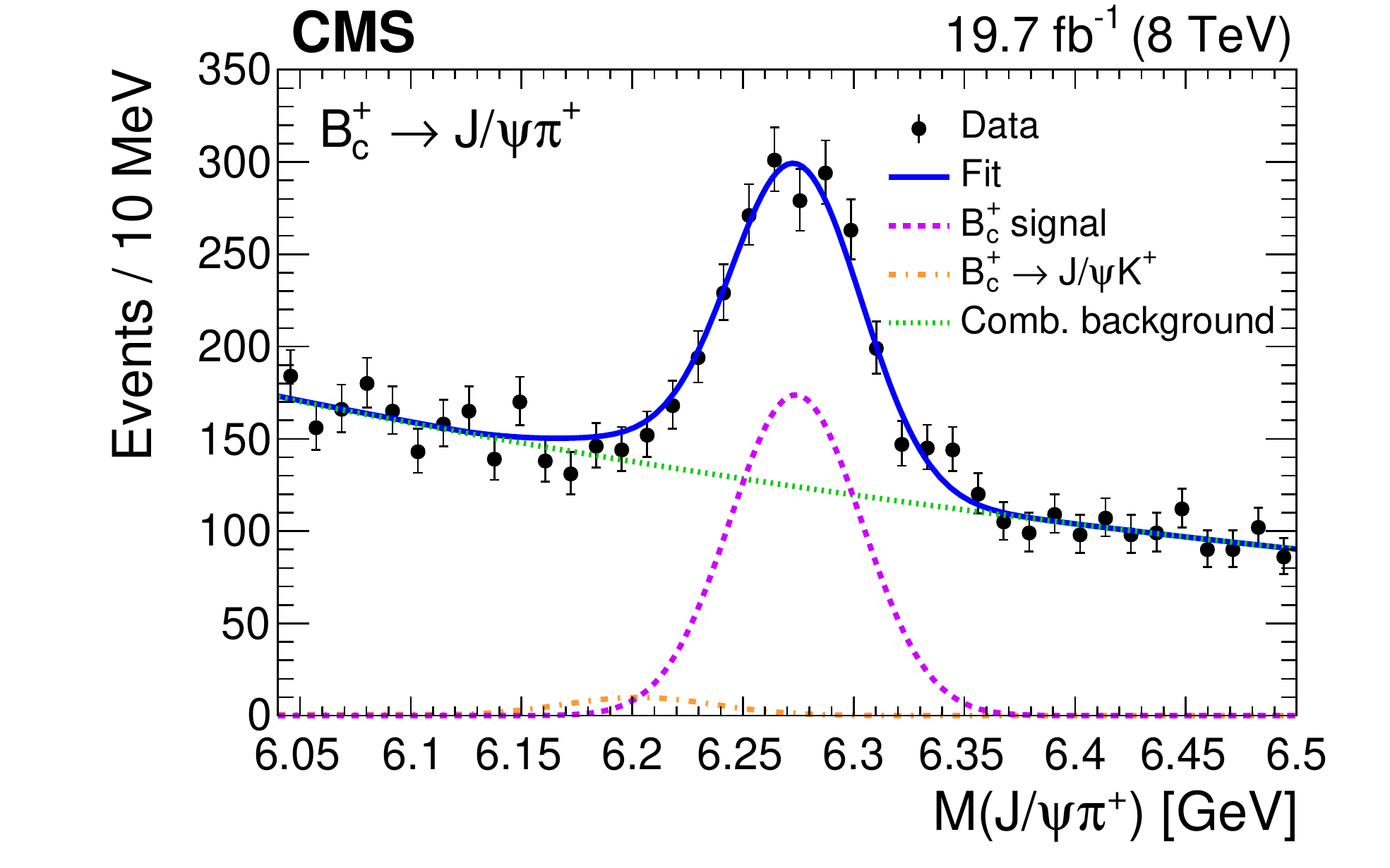}
    \includegraphics[width=0.48\textwidth]{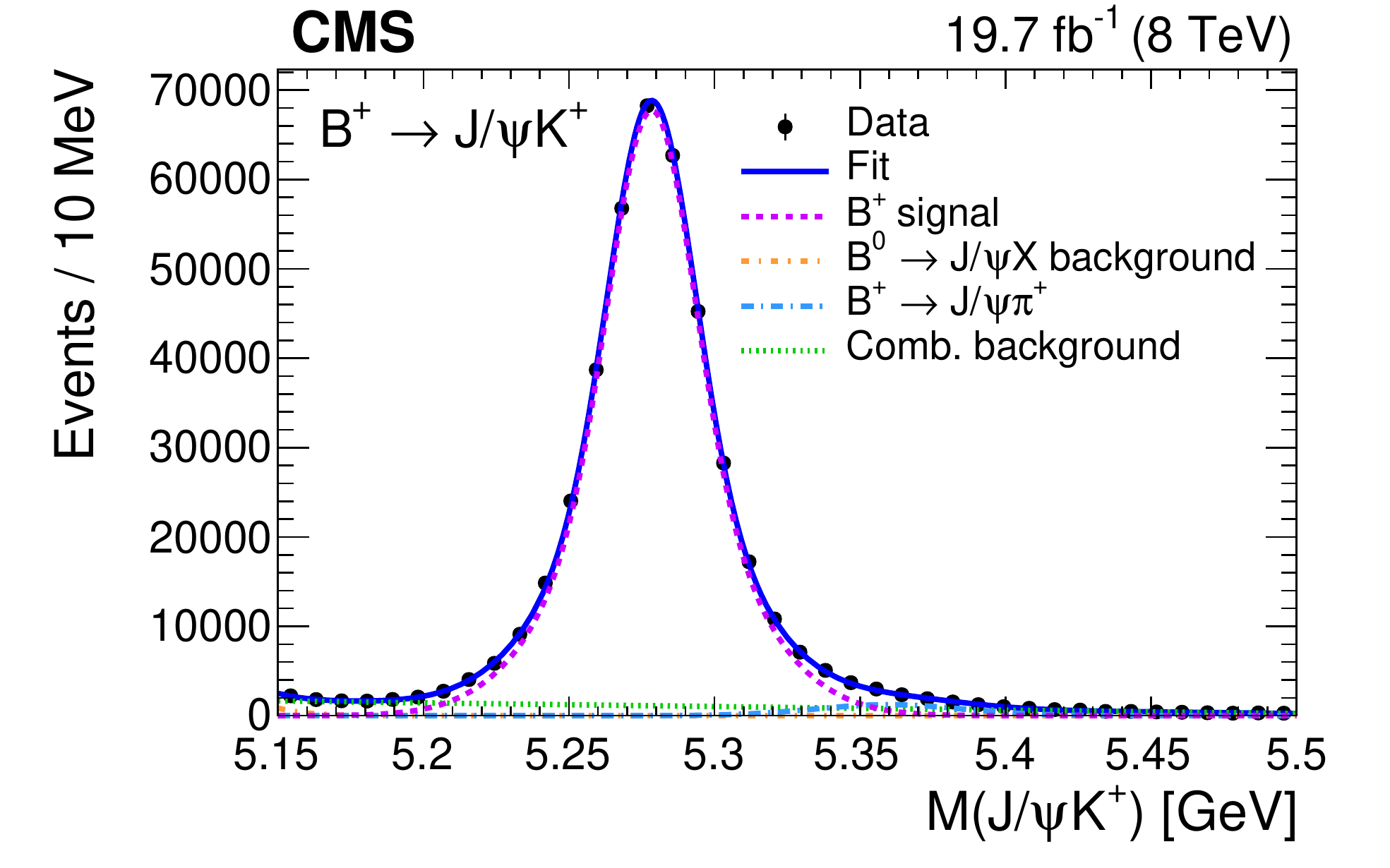}
    \caption{ The \JpsiPi invariant mass distribution (left) with the solid line representing the total fit, the dashed line the signal component, the dotted line the combinatorial background, and the dashed-dotted line the contribution from \BcJpsiKp decays.
    The \JpsiK invariant mass distribution (right) with the solid line representing the total fit, the dashed line the signal component, the dotted-dashed curves the $\PBp \!\to\! \JPsi \Pgpp$ and \PBz{} contributions, and the dotted curve the combinatorial background. The vertical bars on the data points represent the statistical uncertainties.}
    \label{fig:BcBpSignal}
\end{figure}

The \BcJpsiPip and \BJpsiKp invariant mass distributions, shown in Fig.~\ref{fig:BcBpSignal}, are each fit with an unbinned maximum-likelihood estimator.
The \JpsiPi invariant mass distribution is fitted with a Gaussian function for the \PBc signal and an exponential function for the background.
An additional background contribution from \BcJpsiKp decays is modelled from a simulated sample of \BcJpsiKp events, and its contribution is constrained using the value of the branching fraction relative to \JpsiPi~\cite{lhcb_JpsiK}.
The \BcJpsiPip signal yield is $1128 \pm 60$ events, where the uncertainty is statistical only.
The \PBp{} meson invariant mass distribution is fit with a sum of two Gaussian functions with a common mean for the signal and a second-order Chebyshev polynomial for the background.
Additional contributions from partially reconstructed $\PBz$ and \PBp{} meson decays are parametrized with functions determined from  $\PBp\!\to\!\cPJgy \pi^+$ and inclusive $\PBz\!\to\!\cPJgy \mathrm{X}$ MC samples.

\subsection{The fit model and results}

\begin{figure}[hbtp]
\centering
    \includegraphics[width=0.48\textwidth]{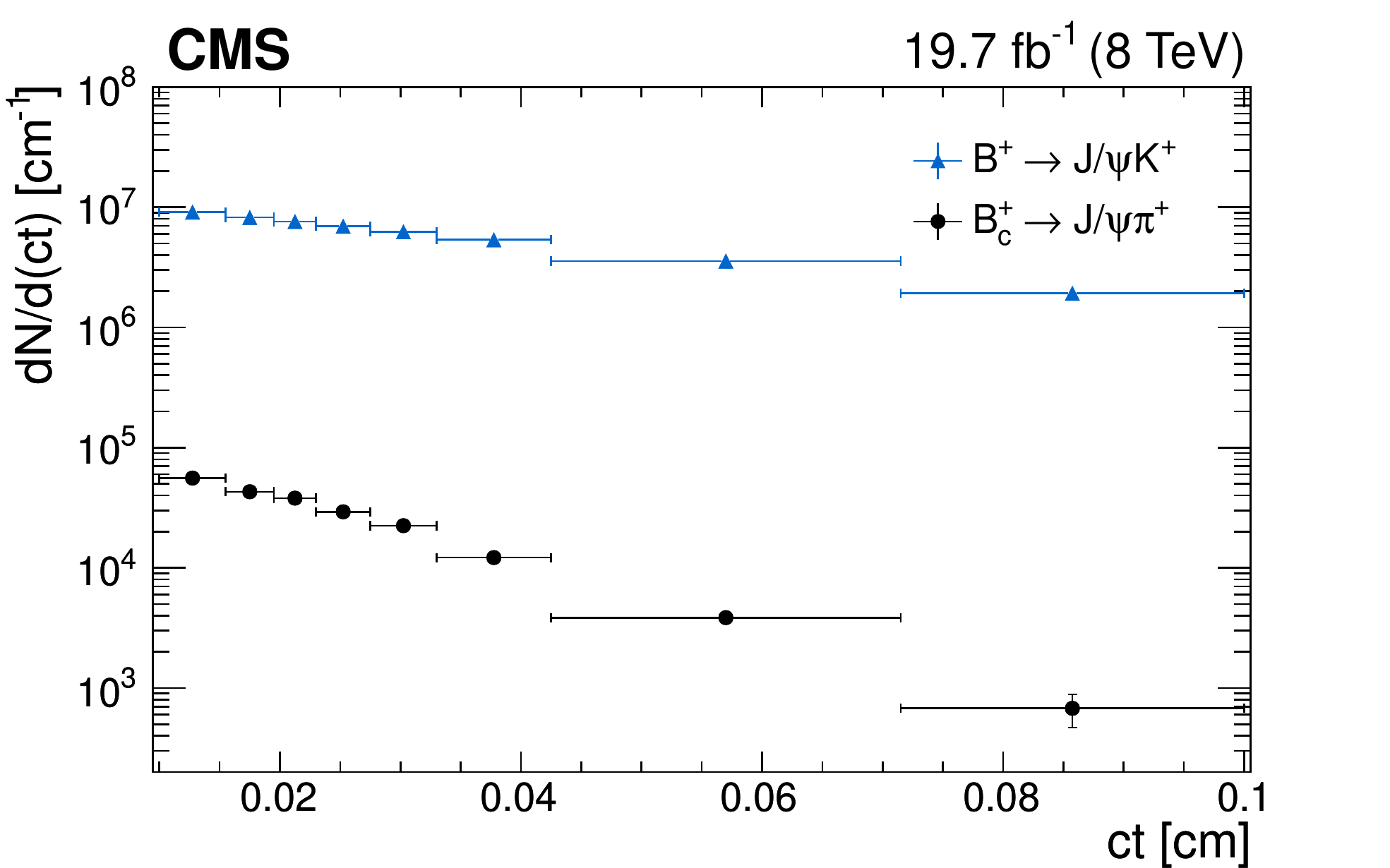}
    \includegraphics[width=0.48\textwidth]{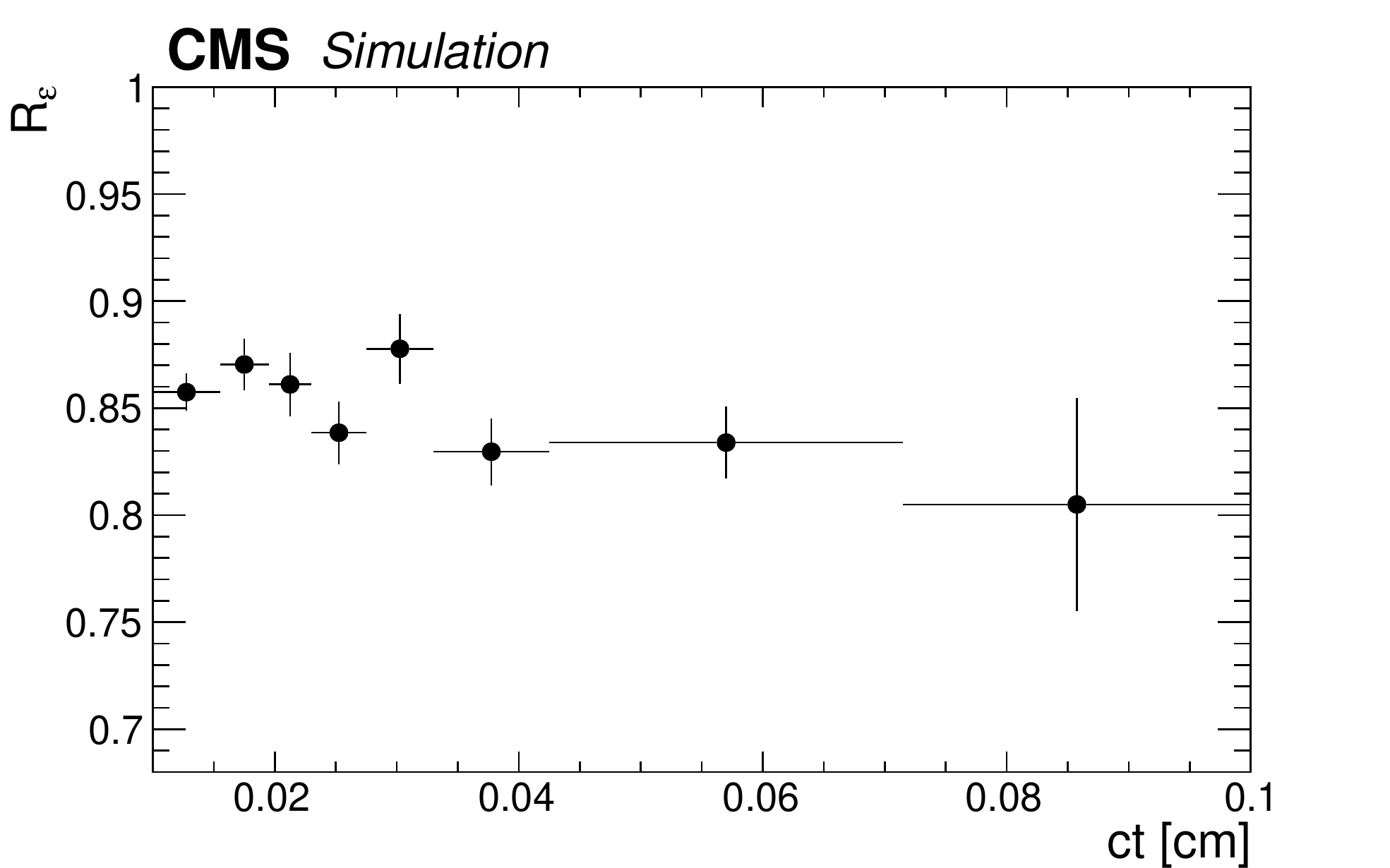}
    \caption{Yields of \BcJpsiPip and \BJpsiKp events (left) as a function of \ct, normalized to the bin width, as determined from fits to the invariant mass distributions. Ratio of the \PBc and \PBp{} efficiency distributions (right) as a function of \ct, as determined from simulated events. The vertical bars on the data points represent the statistical uncertainties, and the horizontal bars show the bin widths.}
    \label{fig:properTimeDistribution_data}
\end{figure}

The \PBc lifetime is extracted through a binned $\chi^2$ fit to the ratio of the efficiency-corrected \ct distributions of the \BcJpsiPip and \BJpsiKp channels.
The \PBc and \PBp{} \ct signal distributions from data are obtained by dividing the data sample into \ct bins and performing an unbinned maximum-likelihood fit to the \JpsiPi and \JpsiK invariant mass distribution in each bin, in the same manner as the fit to the full samples, except that the peak position and resolution are fixed to the values obtained by the fits to the full samples. Varied \ct bin widths are used to ensure a similar statistical uncertainty in the \PBc signal yield among the bins. The bin edges are defined by requiring a relative statistical uncertainty of 12\% or better in each bin.
The same binning is used for the \PBp{} \ct distribution.
The \PBc and \PBp{} meson yields are shown versus \ct in the left plot of Fig.~\ref{fig:properTimeDistribution_data}, where the number of signal events is normalized by the bin width.
Efficiencies are obtained from the MC samples and are defined as the \ct distribution of the selected events after reconstruction divided by the \ct distribution obtained from an exponential decay with the lifetime set to the same value used to generate each MC sample.
The ratio of the two efficiency distributions, using the same binning scheme as for the data, is shown in the right plot of Fig.~\ref{fig:properTimeDistribution_data}.

\begin{figure}[hbtp]
\centering
    \includegraphics[width=0.55\textwidth]{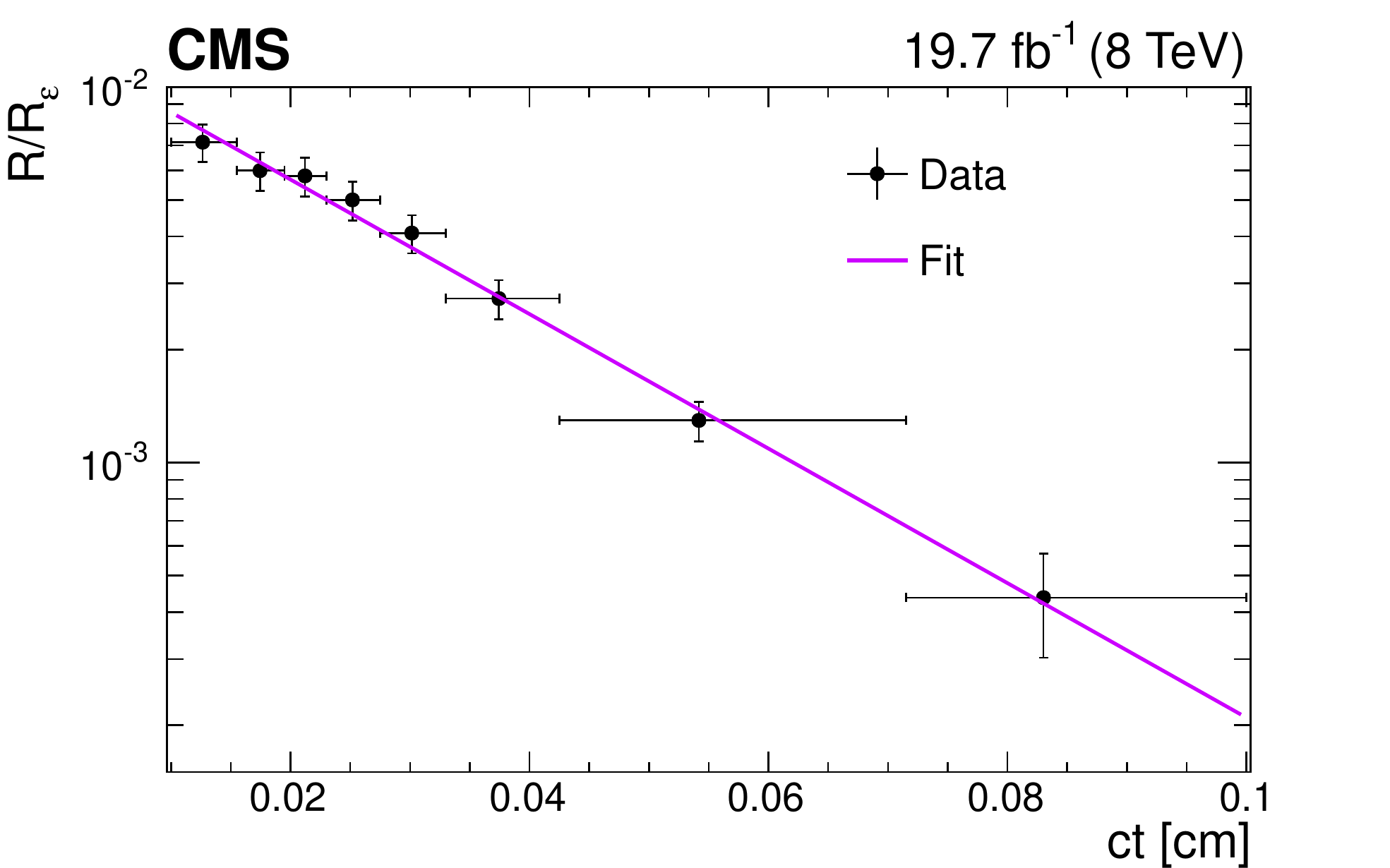}
    \caption{Ratio of the \PBc to \PBp{} efficiency-corrected \ct distributions, $R/R_{\varepsilon}$, with a line showing the result of the fit to an exponential function. The vertical bars give the statistical uncertainty in the data, and the horizontal bars show the bin widths.}
    \label{fig:eff_corr_ratio}
\end{figure}

The ratio of the  \PBc to \PBp{} efficiency-corrected \ct distributions, $R/R_{\varepsilon}$, is shown in Fig.~\ref{fig:eff_corr_ratio}, along with the result of a fit to an exponential function.
The fit returns $\Delta \Gamma = 1.24 \pm 0.09$ ps$^{-1}$.
Using the known lifetime of the \PBp{} meson, $c\tau_{\PBp}$ = 491.1 $\pm$ 1.2 \micron~\cite{Amhis:2016xyh}, a measurement of the \PBc meson lifetime, $c\tau_{\PBc}$ = 162.3 $\pm$ 7.8 \micron, is extracted, where the uncertainty is statistical only.

\section{Systematic uncertainties}\label{sec:syst}

The systematic uncertainties can be divided into uncertainties common to all the measurements, and uncertainties specific to a decay channel. Table~\ref{tab:systcin} summarizes the systematic uncertainties for the sources considered below and the total systematic uncertainty in the \Bs{}, \Bd{}, and \PbgLp{} lifetime measurements.
The systematic uncertainties in $\Delta\Gamma$ and the \PBc meson lifetime are collected in Table~\ref{tab:syst}.
Using the known lifetime of the \PBp{} meson, the uncertainties in $\Delta\Gamma$ are converted into uncertainties in the \PBc meson lifetime measurement. The uncertainty in the \PBc meson lifetime due to the uncertainty in the \PBp{} meson lifetime~\cite{Amhis:2016xyh} is quoted separately.

We have verified that the results are stable against changes in the selection requirements on the quality of the tracks and vertices, the kinematic variables, and \ct, as well as in detector regions and data-taking periods. The effect of replacing the mass of the b hadron in the \ct definition of Eq.~(\ref{eq:ct_formula}) from the world-average to the reconstructed candidate mass is found to be negligible. The lifetimes for all decay channels were measured by treating MC samples as data. No bias was found and all results were consistent with the input lifetimes of the generated samples.

\subsection{Common systematic uncertainties}

\begin{enumerate}
\item Statistical uncertainty in the MC samples\\
The number of events in the simulation directly affects the accuracy of the efficiency determination.
In the case of the \Bs{}, \Bd{}, and \PbgLp{} lifetime measurements, 1000 efficiency curves are generated with variations of the parameter values. The parameter values are sampled using a multivariate Gaussian PDF that is constructed from the covariance matrix of the efficiency fit. The analysis is performed 1000 times, varying the parameters of the efficiency function. The distribution of the measured lifetimes is fitted with a Gaussian function, whose width is taken as the systematic uncertainty associated with the finite size of the simulated samples.
In the measurement of the \PBc lifetime, the bin-by-bin statistical uncertainty in the efficiency determination is propagated to the $R(ct)$ distribution, the fit is performed, and the difference in quadrature of the uncertainty in $\Delta\Gamma$ with respect to the nominal value is taken as the systematic uncertainty.
\item Modelling of the mass distribution shape\\
Biases related to the modelling of the shapes of the b hadron mass signal and background PDFs are quantified by changing the signal and background PDFs individually and using the new models to fit the data. For the \PBz, \PBzs, and \ensuremath{\PgL_\mathrm{b}^0} lifetime measurements, the background model is changed to a higher-degree polynomial, a Chebyshev polynomial, or an exponential function, and the signal model is changed from two Gaussian functions to a single Gaussian function or a sum of three Gaussian functions.
Differences in the measured lifetime between the results of the nominal and alternative models are used to estimate the systematic uncertainty, with the variations due to the modelling of signal and background components evaluated separately and added in quadrature.
For the \PBc lifetime measurement, the signal peak is alternatively modelled with a Crystal Ball distribution~\cite{oreglia}. The alternative description for the background is a first-order Chebyshev distribution. The removal of the Cabibbo-suppressed \BcJpsiKp contribution is also considered.
The maximum deviation of the signal yield in each \ct bin from the nominal value is propagated to the statistical uncertainty in the per-bin yield.
The fit to $R(ct)$ is performed and the difference in quadrature between the uncertainty from this fit and the nominal measurement is taken as the systematic uncertainty.
\end{enumerate}

\subsection{Channel-specific systematic uncertainties}

\begin{enumerate}
\item Modelling of the background \ct shape in the \Bs{}, \Bd{}, and \PbgLp{} channels\\
To estimate a systematic uncertainty due to the \ct background model, we add an additional background contribution modelled with its own lifetime, and
compare the result to that obtained with the nominal fit model. The difference between the results of the nominal and alternative fit models is used as the systematic uncertainty from the \ct shape modelling.
\item The \Bu{} contamination in the \BsJpsiPiPi{} sample\\
In the nominal fit, the yield and lifetime of the \BJpsiKp contamination are determined from the fit with the mass shape obtained from simulation. An alternative estimate of the \JpsiK{} contamination is obtained from data by taking the leading pion of the \BsJpsiPiPi{} decay to be the kaon.  The lifetime and yield of the \BJpsiKp decays contaminating the \BsJpsiPiPi{} sample are determined from a fit of the \Bu{} signal candidates in the \BsJpsiPiPi{} sample, with the mass shape also obtained from the data. The difference between the \Bs{} lifetime found with this model and the nominal model is considered as the systematic uncertainty due to \Bu{} contamination.
\item Invariant mass window of the $\pi^+ \pi^-$ in the \BsJpsiPiPi{} channel\\
Although the events selected by the $\pi^+\pi^-$ mass window are dominated by the $f_0(980)$, its width is not well known and possible backgrounds under the $f_0(980)$ peak could be increased or decreased, depending on the mass window. The effect on the lifetime is studied by using mass windows of $\pm$30 and $\pm$80\MeV around the signal peak, compared to the nominal fit result with a
$\pm$50\MeV window.  The maximum variation of the lifetime is taken as the systematic uncertainty.
\item The $\PKp\Pgpm$ mass assignments for \Kstar{} candidates in the \BdJpsiKstar{} channel\\
The \Kstar{} candidates are constructed from a pair of tracks with kaon and pion mass assignments. The combination with invariant mass closest to the world-average \Kstar{} mass is chosen to reconstruct the \Bd{} candidate. To estimate the effect on the lifetime due to a possible misassignment of kaon and pion, both combinations are discarded if both are within the natural width of the \Kstar{} mass, and the difference between the lifetime obtained with this sample and the nominal
sample is taken as the systematic uncertainty.
\item The \ct range in the \BsJpsiPhi{} channel\\
Since the $ct>0.02$ cm requirement distorts the fractions of heavy and light mass eigenstates, the measured \Bs{} effective lifetime must be corrected. The correction and systematic uncertainty are quantified analytically. The correction to the effective lifetime is
\ifthenelse{\boolean{cms@external}}{
 \begin{multline}
\delta_{ct} =
c\tau_{\text{eff}}^{\mathrm{cut}} - c\tau_{\text{eff}} = \\
\frac{ (1-|A_\perp|^2) (c\tau_{\mathrm{L}})^2 \re^{-a/c\tau_\mathrm{L}} + |A_\perp|^2 (c\tau_{\mathrm{H}})^2  \re^{-a/c\tau_\mathrm{H}} }{(1-|A_\perp|^2) c\tau_{\mathrm{L}} \re^{-a/c\tau_\mathrm{L}} + |A_\perp|^2 c\tau_{\mathrm{H}}  \re^{-a/c\tau_\mathrm{H}} } \\- \frac{ (1-|A_\perp|^2) (c\tau_{\mathrm{L}})^2 + |A_\perp|^2 (c\tau_{\mathrm{H}})^2 }{ (1-|A_\perp|^2) c\tau_{\mathrm{L}}+ |A_\perp|^2 c\tau_{\mathrm{H}} } ,
\label{eq:PDL_correction}
\end{multline}
}{
\begin{equation}
 \begin{split}
\delta_{ct} = c\tau_{\text{eff}}^{\mathrm{cut}} - c\tau_{\text{eff}} = \frac{ (1-|A_\perp|^2) (c\tau_{\mathrm{L}})^2 \re^{-a/c\tau_\mathrm{L}} + |A_\perp|^2 (c\tau_{\mathrm{H}})^2  \re^{-a/c\tau_\mathrm{H}} }{(1-|A_\perp|^2) c\tau_{\mathrm{L}} \re^{-a/c\tau_\mathrm{L}} + |A_\perp|^2 c\tau_{\mathrm{H}}  \re^{-a/c\tau_\mathrm{H}} } \\- \frac{ (1-|A_\perp|^2) (c\tau_{\mathrm{L}})^2 + |A_\perp|^2 (c\tau_{\mathrm{H}})^2 }{ (1-|A_\perp|^2) c\tau_{\mathrm{L}}+ |A_\perp|^2 c\tau_{\mathrm{H}} } ,
 \end{split}
\label{eq:PDL_correction}
\end{equation}
}
where the first term represents the effective lifetime in the presence of a $ct>a$ requirement and the latter term is the unbiased effective lifetime. In this analysis, $a$ is equal to 0.02\cm.
The world-average values~\cite{PDG_lifetimes16} for $c\tau_\mathrm{H}=482.7 \pm 3.6\micron$, $c\tau_\mathrm{L}= 426.3 \pm 2.4\micron$, and ${|A_\perp|^2 = 0.250 \pm 0.006}$ are used to obtain the correction  ${\delta_{ct} = 0.62 \pm 0.10\micron}$.
\item The S-wave contamination in the \BsJpsiPhi{} channel\\
The \Bs{} candidates reconstructed in the \JpsiPhi{} final state contain a small fraction of nonresonant and CP-odd \BsJpsiKK{} decays, where the invariant mass of the two kaons happens to be near the $\phi$ meson mass.
The fraction of \BsJpsiKK{} decays among the selected events is measured in the weak mixing phase analysis~\cite{phisPaper} to be $f_\mathrm{S} = (1.2^{+0.9}_{-0.7})$\%. Because of the different trigger and signal selection criteria of the present analysis, the S-wave fraction is corrected according to the simulation to be $(1.5^{+1.1}_{-0.9})$\%.
The bias caused by the contamination of nonresonant \BsJpsiKK{} decays is estimated by generating two sets of pseudo-experiments, one with just \BsJpsiPhi{} events and one with a fraction of S-wave events based on the measured S-wave fraction and its uncertainty.  The difference in the average of the measured lifetimes of these two samples is 0.74\micron, which is used to correct the measured lifetime.  The systematic uncertainty associated with this correction is obtained by taking the difference in quadrature between the standard deviation of the distribution of lifetime results from the pseudo-experiments with and without the S-wave contribution.
\item PV selection in the \BcJpsiPip{} channel\\
From the multiple reconstructed PVs in an event, one is selected to compute the \ct{} value of the candidate.  Two alternative methods to select the PV position are studied: using the centre of the beamspot and selecting the PV with the largest sum of track \pt.  While all three methods are found to be effective and unbiased, there were small differences, and the maximum deviation with respect to the nominal choice is taken as the systematic uncertainty.  The \Bu{} and \PBc{} primary vertex choices were changed coherently.
\item Detector alignment in the \BcJpsiPip channel\\
Possible effects on the lifetime due to uncertainties in the detector alignment~\cite{CMSaligment} are investigated for each decay topology using 20 different simulated samples with distorted geometries.  These distortions include expansions in the radial and longitudinal dimensions, rotations, twists, offsets, etc.  The amount of misalignment is chosen such that it is large enough to be detected and corrected by the alignment procedure.  The standard deviation of the lifetimes for the tested scenarios is taken as the systematic uncertainty from this source.  The \Bu{} and \PBc{} geometries were changed coherently.
\item Absolute $ct$ accuracy in the \Bs{}, \Bd{}, and \PbgLp{} lifetime measurements\\
The lifetime of the most statistically precise mode (\BpJpsiKp) is used to validate the accuracy of the simulation and various detector calibrations. The difference between our measurement of $490.9 \pm 0.8\micron$ (statistical uncertainty only) and the world-average value of $491.1 \pm 1.2\micron$~\cite{Amhis:2016xyh} is $0.2 \pm 1.4\micron$.  This implies a limit to the validation of $1.4/491 = 0.3\%$.  Four systematic effects that we expect to be included were checked independently.  The systematic uncertainties from PV selection and detector alignment were found to be 0.7\micron and 0.3--0.7\micron, respectively.  Varying the efficiency functional form changed the lifetimes by 0.3--0.6\micron, while varying $\sigma_{ct}$ by factors of 0.5 and 2.0 resulted in lifetime differences of no more than $0.2\micron$.  As the sum in quadrature of these uncertainties is less than that obtained from the \Bu{} lifetime comparison, we assign a value of 0.3\% as the systematic uncertainty for the absolute $ct$ accuracy.
\end{enumerate}

\begin{table*} [h!tbp]
\centering
\topcaption{Summary of the sources and values of systematic uncertainties in the lifetime measurements (in \micron).  The total systematic uncertainty is the sum in quadrature of the individual uncertainties. \label{tab:systcin}}
\begin{tabular}[t]{@{}l@{\,~}|@{\,~}c@{~~}c@{~~}c@{~~}c@{~~}c@{~~}c@{}}
 Source            & $\Bd \!\!\to\! \Jpsinarrow\PKstnarrow{}^{\hspace{-0.04em}0} $   & $\Bd \!\!\to\! \Jpsinarrow\PKzS$ & $\Bs \!\!\to\! \Jpsinarrow\pi^+\pi^-$ & $\Bs \!\!\to\! \Jpsinarrow\phi$ & $\PbgLp \!\!\to\! \Jpsinarrow{} \Lambda^{\!0}$  \\
\hline
MC statistical uncertainty              &  1.1           &   2.4                      &    2.0                       &       0.6     & 2.3    \\
Mass modelling                          &  0.3           &   0.4                      &    0.2                       &       0.4     & 0.9    \\
\ct modelling                          &  0.1           &   0.1                      &    0.4                       &       0.0     & 0.1    \\
\Bu{} contamination                     &  ---           &   ---                      &    1.4                       &       ---     & ---    \\
Mass window of $\pi^+ \pi^-$            &  ---           &   ---                      &    1.8                       &       ---     & ---    \\
$\PK^\pm\pi^\mp$ mass assignment &  $0.3$         &   ---                      &    ---                       &       ---     & ---    \\
\ct range                              &  ---           &   ---                      &    ---                       &       0.1     & ---    \\
S-wave contamination                    &  ---           &   ---                      &    ---                       &       0.4     & ---    \\
Absolute $ct$ accuracy                   &  1.3           &   1.3                      &    1.4                       &       1.3     & 1.3    \\
\hline
Total ($\mu$m)                                 	&  1.8           & 2.8                    & 3.4                        &  1.5          & 2.8  \\
\end{tabular}
\end{table*}

\begin{table}[ht]
\caption{Summary of the systematic uncertainties in the $\Delta\Gamma$ and $c\tau_{\PBc}$ measurements.}
\label{tab:syst}
\centering
\begin{tabular}{l|c|c}
Source       			  & $\Delta\Gamma$ [ps$^{-1}$]    & $c\tau_{\PBc}$ [$\mu\text{m}$]   \\[4pt] \hline
MC statistical uncertainty& 0.01		&	1.2 \\
Mass modelling  	      & 0.04	    &	3.4 \\
PV selection              & 0.02		&   2.0 \\
Detector alignment		  & 0.01	    & 	0.6 \\	
\hline
Total uncertainty		  & 0.05	    &   4.2 \\
\end{tabular}
\end{table}

\section{Lifetime measurement results}

Our final results for the \Bd{}, \Bs{}, and \PbgLp{} hadron lifetimes are:
\begin{align}
 c\tau_{\PBz \to \cPJgy \Kstar}    &=   453.0 \pm 1.6\stat \pm 1.8\syst\mum,\\
 c\tau_{\PBz \to \cPJgy \PKzS}    &=   457.8 \pm 2.7\stat \pm 2.8\syst\mum,\\
 c\tau_{\PBzs \to \cPJgy \Pgpp \Pgpm}  &=   502.7 \pm 10.2\stat \pm 3.4\syst\mum, \label{eq:JpsiPiPiresult}\\
 c\tau_{\PBzs \to \cPJgy \Pgf}   &=   443.9  \pm 2.0\stat \pm 1.5\syst\mum,\label{eq:JpsiPhiresult}\\
 c\tau_{\PbgLp{}}  &=   442.9  \pm 8.2\stat \pm 2.8\syst\mum.
 \end{align}
The value of the \Bs{} lifetime using the \JpsiPhi{}  decay has been corrected for the \ct{} range and S-wave contamination effects described in Section~\ref{sec:syst}.
The lifetime ratios $\tau_{\Bs}/\tau_{\Bd}$ and $\tau_{\PbgLp}/\tau_{\Bd}$ have been determined using the decay channels
\BdJpsiKstar, \BsJpsiPhi, and \LambJpsiLam.  Including the statistical and correlated and uncorrelated systematic uncertainties, the results are:
\ifthenelse{\boolean{cms@external}}{
\begin{multline}
\tau_{\PbgLp}/\tau_{\PBz \to \cPJgy \Kstar} \\= 0.978 \pm 0.018\stat \pm 0.006\syst,  
\end{multline}
\begin{multline}
\tau_{\PBzs \to \cPJgy \Pgf}/\tau_{\PBz \to \cPJgy \Kstar}  \\= 0.980 \pm 0.006\stat \pm 0.003\syst.
\end{multline}}
{
\begin{align}
\tau_{\PbgLp}/\tau_{\PBz \to \cPJgy \Kstar} &= 0.978 \pm 0.018\stat \pm 0.006\syst,  \\
\tau_{\PBzs \to \cPJgy \Pgf}/\tau_{\PBz \to \cPJgy \Kstar}  &= 0.980 \pm 0.006\stat \pm 0.003\syst.
\end{align}
}
These ratios are compatible with the current world-average values.

The measured lifetimes for the \PBz\ meson in the two different channels are in agreement.
Combining the two results, including the statistical and the correlated and uncorrelated systematic uncertainties,
gives $c\tau_{\PBz} = 454.1 \pm 1.4\stat \pm 1.7\syst\micron$.
The lifetime measurements can also be used to estimate $\Gamma_\mathrm{d}$ and $\Delta \Gamma_\mathrm{d}$~\cite{DGtheory}.
In the standard model, the effective lifetimes of the two \PBz\ decay modes can be written as:
\begin{align}
\tau_{\PBz \to \cPJgy \PKzS} &= \frac{1}{\Gamma_\mathrm{d}} \left(\frac{1}{1-y_\mathrm{d}^2}\right)  \left( \frac{1+2\cos{(2\beta)}y_\mathrm{d}+y_\mathrm{d}^2}{1+\cos{(2\beta)}y_\mathrm{d}}\right) ,  \\
\tau_{\PBz \to \cPJgy \PKst{}^{0}} &= \frac{1}{\Gamma_\mathrm{d}}  \left(\frac{1+y_\mathrm{d}^2}{1-y_\mathrm{d}^2}\right),
\end{align}
where $y_\mathrm{d}=\Delta \Gamma_\mathrm{d}/2\Gamma_\mathrm{d}$, and $\beta=(21.9\pm0.7)^\circ$~\cite{Amhis:2016xyh} is one of the CKM unitarity triangle angles.
Using our measured values for the two \PBz\ lifetimes, we fit for $\Gamma_\mathrm{d}$ and $\Delta \Gamma_\mathrm{d}$ and use the values to
determine $\Delta \Gamma_\mathrm{d}/\Gamma_\mathrm{d}$.  The results are:
\begin{align}
\Gamma_\mathrm{d} &= 0.662 \pm 0.003\stat \pm 0.003\syst\,\mathrm{ps}^{-1},  \\
\Delta \Gamma_\mathrm{d} &=  0.023 \pm 0.015\stat \pm 0.016\syst\,\mathrm{ps}^{-1}, \\
\Delta \Gamma_\mathrm{d}/\Gamma_\mathrm{d} &= 0.034 \pm 0.023\stat \pm 0.024\syst.
\end{align}

Neglecting CP violation in mixing, the measured \BsJpsiPiPi{}  lifetime can be translated into the width of the heavy \Bs{} mass eigenstate:
\begin{equation}
    \Gamma_\mathrm{H} = 1/\tau_{\Bs} = 0.596 \pm 0.012\stat \pm 0.004\syst\,\mathrm{ps}^{-1}.
\end{equation}
Solving for $c\tau_{\mathrm{L}}$ from Eq.~(\ref{eq:Bs_tau_eff}) gives
\begin{equation}
c\tau_\mathrm{L} = \frac{1}{2} c\tau_{\text{eff}} + \sqrt{ \frac{1}{4} (c\tau_{\text{eff}})^2 - \frac{|A_\perp|^2}{1-|A_\perp|^2} c\tau_\mathrm{H}( c\tau_\mathrm{H} - c\tau_{\text{eff}})}.
\end{equation}
Using the \BsJpsiPiPi{} result in Eq.~(\ref{eq:JpsiPiPiresult}), the measured \Bs{} effective lifetime in Eq.~(\ref{eq:JpsiPhiresult}), and the world-average value of the magnitude squared of the CP-odd amplitude $|A_\perp|^2 = 0.250 \pm 0.006$~\cite{PDG_lifetimes16}, the lifetime of the light component is found to be $c\tau_{\mathrm{L}} = 420.4 \pm 6.2\micron$. The uncertainty includes all statistical and systematic uncertainties, taking into account the correlated uncertainties. The result is consistent with the world-average value of $423.6 \pm 1.8\micron$~\cite{Amhis:2016xyh}.

Our measured lifetimes for \PBz, \BsJpsiPhi, and \PbgLp\ are compatible with the
current world-average values~\cite{Amhis:2016xyh} of $455.7 \pm 1.2$\micron,
$443.4 \pm 3.6$\micron, and $440.7 \pm 3.0$\micron, respectively. In addition, our
measurement of the \Bs\ lifetime in the \BsJpsiPiPi\ channel is in agreement with
the results from CDF, LHCb, and D0: $510\,^{+36}_{-33}\stat \pm 9\syst$\micron~\cite{BspipiLifetimeCDF},
$495.3 \pm 7.2\stat \pm 7.2\syst$\micron~\cite{BspipiLifetimeLHCB}, and
$508 \pm 42\stat\pm 16\syst$\micron~\cite{BspipiLifetimeD0}, respectively.

Our final result for the \PBc{} lifetime using the \JpsiPi{} mode is:
\begin{equation}
  c\tau_{\PBc} = 162.3 \pm 7.8\stat \pm 4.2\syst \pm 0.1  (\tau_{\PBp})\mum,
\end{equation}
where the systematic uncertainty from the \PBp{} lifetime uncertainty~\cite{Amhis:2016xyh} is quoted separately in the result.
This measurement is in agreement with the world-average value $(152.0 \pm 2.7\micron)$~\cite{Amhis:2016xyh}.
Precise measurements of the \PBc lifetime allow tests of various theoretical models, which predict values ranging from 90 to 210\micron~\cite{Chang:2000ac,Beneke:1996xe,Anisimov:1998uk,Kiselev:2000nf}.
Furthermore, they provide new constraints on possible physics beyond the standard model from the observed anomalies in $\PB \to \PD^{(*)}\tau\nu$ decays~\cite{Alonso:2016oyd}.

\section{Summary}

The lifetimes of the \PBz, \PBzs, \PbgLp, and \PBc{} hadrons have been measured using fully reconstructed decays with a \Jpsi{} meson.
The data were collected by the CMS detector in proton-proton collision events at a centre-of-mass energy of 8\TeV,
and correspond to an integrated luminosity of 19.7\fbinv.
The \PBz{} and \PBzs{} meson lifetimes have each been measured in two channels: \JpsiKstar, \JpsiKshort\ for \PBz\ and\JpsiPiPi, \JpsiPhi\ for \PBzs.
The precision from each channel is as good as or better than previous measurements in the respective channel.
The \PBz{} lifetime results are used to obtain an average lifetime and to measure the decay width difference between the two mass eigenstates.
The \PBzs\ lifetime results are used to obtain the lifetimes of the heavy and light \Bs\ mass eigenstates.
The precision of the \PbgLp\ lifetime measurement is also as good as any previous measurement in the \JpsiLam\ channel.
The measured \PBc\ meson lifetime is in agreement with the results from LHCb and significantly more precise than the CDF and D0 measurements.
All measured lifetimes are compatible with the current world-average values.

\begin{acknowledgments}

We congratulate our colleagues in the CERN accelerator departments for the excellent performance of the LHC and thank the technical and administrative staffs at CERN and at other CMS institutes for their contributions to the success of the CMS effort. In addition, we gratefully acknowledge the computing centres and personnel of the Worldwide LHC Computing Grid for delivering so effectively the computing infrastructure essential to our analyses. Finally, we acknowledge the enduring support for the construction and operation of the LHC and the CMS detector provided by the following funding agencies: BMWFW and FWF (Austria); FNRS and FWO (Belgium); CNPq, CAPES, FAPERJ, and FAPESP (Brazil); MES (Bulgaria); CERN; CAS, MoST, and NSFC (China); COLCIENCIAS (Colombia); MSES and CSF (Croatia); RPF (Cyprus); SENESCYT (Ecuador); MoER, ERC IUT, and ERDF (Estonia); Academy of Finland, MEC, and HIP (Finland); CEA and CNRS/IN2P3 (France); BMBF, DFG, and HGF (Germany); GSRT (Greece); OTKA and NIH (Hungary); DAE and DST (India); IPM (Iran); SFI (Ireland); INFN (Italy); MSIP and NRF (Republic of Korea); LAS (Lithuania); MOE and UM (Malaysia); BUAP, CINVESTAV, CONACYT, LNS, SEP, and UASLP-FAI (Mexico); MBIE (New Zealand); PAEC (Pakistan); MSHE and NSC (Poland); FCT (Portugal); JINR (Dubna); MON, RosAtom, RAS, RFBR and RAEP (Russia); MESTD (Serbia); SEIDI, CPAN, PCTI and FEDER (Spain); Swiss Funding Agencies (Switzerland); MST (Taipei); ThEPCenter, IPST, STAR, and NSTDA (Thailand); TUBITAK and TAEK (Turkey); NASU and SFFR (Ukraine); STFC (United Kingdom); DOE and NSF (USA).

\hyphenation{Rachada-pisek} Individuals have received support from the Marie-Curie programme and the European Research Council and Horizon 2020 Grant, contract No. 675440 (European Union); the Leventis Foundation; the A. P. Sloan Foundation; the Alexander von Humboldt Foundation; the Belgian Federal Science Policy Office; the Fonds pour la Formation \`a la Recherche dans l'Industrie et dans l'Agriculture (FRIA-Belgium); the Agentschap voor Innovatie door Wetenschap en Technologie (IWT-Belgium); the Ministry of Education, Youth and Sports (MEYS) of the Czech Republic; the Council of Science and Industrial Research, India; the HOMING PLUS program of the Foundation for Polish Science, cofinanced from European Union, Regional Development Fund, the Mobility Plus program of the Ministry of Science and Higher Education, the National Science Center (Poland), contracts Harmonia 2014/14/M/ST2/00428, Opus 2014/13/B/ST2/02543, 2014/15/B/ST2/03998, and 2015/19/B/ST2/02861, Sonata-bis 2012/07/E/ST2/01406; the National Priorities Research Program by Qatar National Research Fund; the Programa Severo Ochoa del Principado de Asturias; the Thalis and Aristeia programs cofinanced by EU-ESF and the Greek NSRF; the Rachadapisek Sompot Fund for Postdoctoral Fellowship, Chulalongkorn University and the Chulalongkorn Academic into Its 2nd Century Project Advancement Project (Thailand); the Welch Foundation, contract C-1845; and the Weston Havens Foundation (USA).
\end{acknowledgments}

\bibliography{auto_generated}

\cleardoublepage \appendix\section{The CMS Collaboration \label{app:collab}}\begin{sloppypar}\hyphenpenalty=5000\widowpenalty=500\clubpenalty=5000\vskip\cmsinstskip
\textbf{Yerevan Physics Institute,  Yerevan,  Armenia}\\*[0pt]
A.M.~Sirunyan,  A.~Tumasyan
\vskip\cmsinstskip
\textbf{Institut f\"{u}r Hochenergiephysik,  Wien,  Austria}\\*[0pt]
W.~Adam,  F.~Ambrogi,  E.~Asilar,  T.~Bergauer,  J.~Brandstetter,  E.~Brondolin,  M.~Dragicevic,  J.~Er\"{o},  M.~Flechl,  M.~Friedl,  R.~Fr\"{u}hwirth\cmsAuthorMark{1},  V.M.~Ghete,  J.~Grossmann,  J.~Hrubec,  M.~Jeitler\cmsAuthorMark{1},  A.~K\"{o}nig,  N.~Krammer,  I.~Kr\"{a}tschmer,  D.~Liko,  T.~Madlener,  I.~Mikulec,  E.~Pree,  N.~Rad,  H.~Rohringer,  J.~Schieck\cmsAuthorMark{1},  R.~Sch\"{o}fbeck,  M.~Spanring,  D.~Spitzbart,  W.~Waltenberger,  J.~Wittmann,  C.-E.~Wulz\cmsAuthorMark{1},  M.~Zarucki
% ...
\vskip\cmsinstskip
% ...
1:  Also at Vienna University of Technology,  Vienna,  Austria\\
% ...
\end{sloppypar}
\end{document}